\documentclass[prd, preprint, longbibliography, 11pt]{revtex4-1}

\usepackage{amsmath}
\usepackage{mathtools}
\usepackage{amssymb}
\usepackage{setspace}
\usepackage{graphicx}
\usepackage{float}
\usepackage{tensor}
\usepackage[utf8]{inputenc}
\usepackage{amsfonts}
\usepackage{braket}
\usepackage{esint}
\usepackage{breqn}
\usepackage{IEEEtrantools}
\usepackage{hyperref}
\usepackage{caption}
\usepackage{soul}
\usepackage{ragged2e}

\begin{document}
 
 %

\begin{center}
{ \large \bf The Life and Science of Thanu Padmanabhan (1957-2021)}

\smallskip


{\large{{\it remembered} by \bf The Paddy Gharana}, {\it with contributions from}}


\justify{Jasjeet Singh Bagla, Krishnakanta Bhattacharya, Sumanta Chakraborty, Sunu Engineer, Valerio Faraoni,  Sanved Kolekar, Dawood Kothawala, Kinjalk Lochan, Sujoy Modak,  V. Parameswaran Nair, Aseem Paranjape, Krishnamohan Parattu, Sarada G. Rajeev, Bibhas Ranjan Majhi, Tirthankar Roy Choudhury, Mohammad Sami, Sudipta Sarkar, Sandipan Sengupta, T. R. Seshadri, S. Shankaranarayanan, Suprit Singh, Tejinder P. Singh, L. Sriramkumar and Urjit  Yajnik}

\smallskip

\end{center}

\smallskip
\begin{figure*}[!h]
\captionsetup{labelformat=empty,listformat=empty}
\centering
\includegraphics[width=12.00cm]{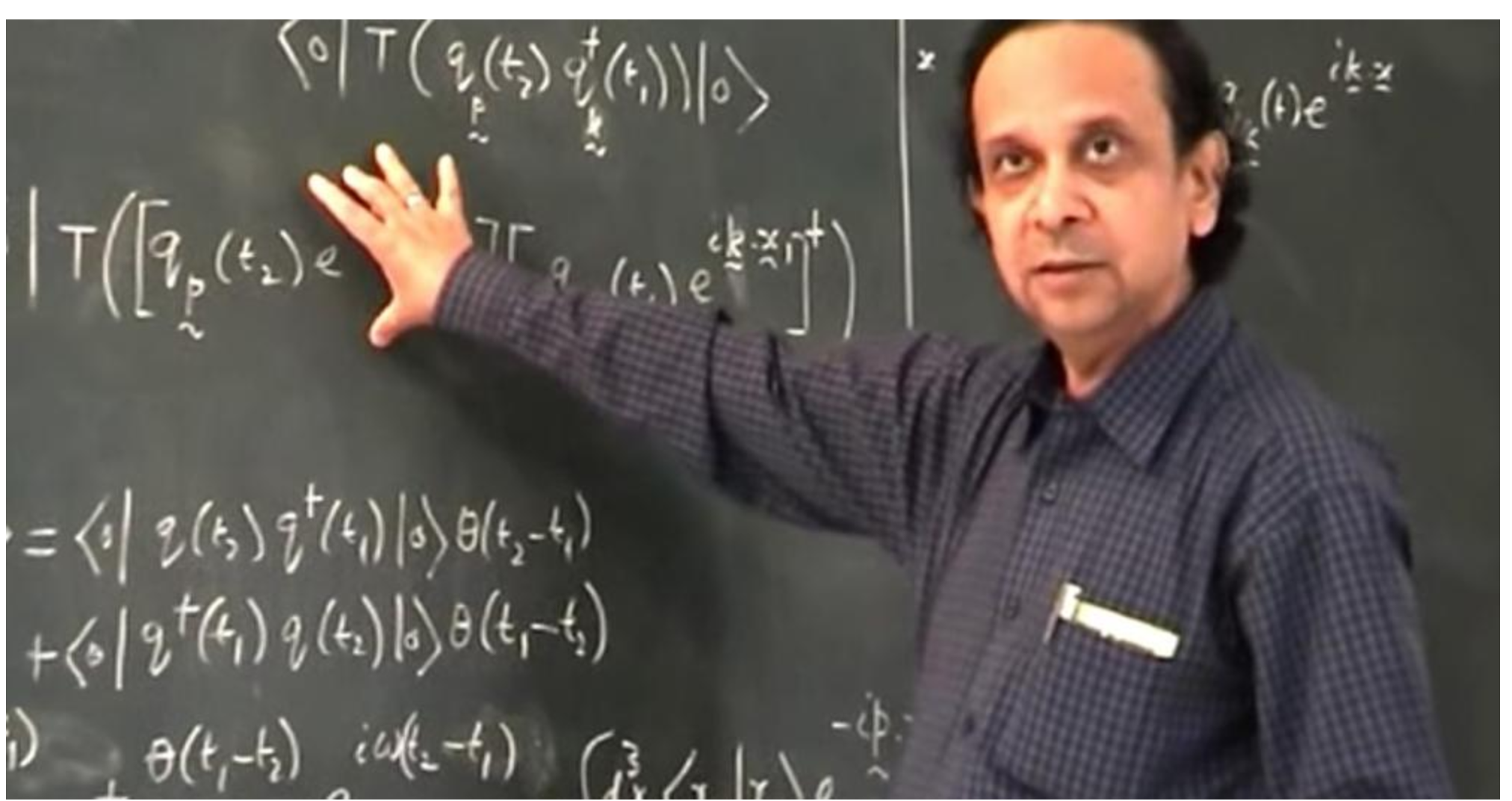}
\caption{\url{https://www.iucaa.in/~paddy/}} 
\end{figure*}

\centerline{\bf ABSTRACT}
\noindent Thanu Padmanabhan was a renowned Indian theoretical physicist known for his research in general relativity, cosmology, and quantum gravity. In an extraordinary career spanning forty-two years, he published more than three hundred research articles, wrote ten highly successful technical and popular books, and mentored nearly thirty graduate students and post-doctoral fellows. He is best known for his deep work investigating gravitation as an emergent thermodynamic phenomenon. He was an outstanding teacher, and an indefatigable populariser of science, who travelled very widely to motivate and inspire young students. Paddy, as he was affectionately known, was also a close friend to his students and collaborators, treating them as part of his extended academic family. On September 17, 2021 Paddy passed away very unexpectedly, at the age of sixty-four and at the height of his research career, while serving as a Distinguished Professor at the Inter-University Centre for Astronomy and Astrophysics, Pune. His untimely demise has come as a shock to his family and friends and colleagues. In this article, several  of them have come together to pay their tributes and share their fond memories of Paddy.

\newpage

\section{Professor Padmanabhan: A brief bio-sketch}

\noindent {\it  Jasjeet Singh Bagla and Sunu Engineer}

\noindent  Thanu Padmanabhan (hereafter Paddy)  was born on 10 March 1957 in a
lower middle class family at Trivandrum, Kerala.  
His mother, Lakshmi, was a home-maker. His father, Thanu Iyer, had  a
genius for mathematics but had to abandon his academic pursuits
because of family circumstances and take up a job in the forest
department of the government of Kerala.
However, his father, as well as several other family members of his
father's generation,  had a great passion for all of mathematics,
especially geometry.  
Two strong inspirations in Padmanabhan's early life, which influenced
him to take up academic pursuits, were his father and another senior
member of the family, Neelakanta Sarma.
Both of them had a high level of personal integrity and passion for
knowledge --- two qualities which Paddy inherited  with  the code of
life emphasised in the family circles in which he grew up being:
``\textit{Excellence is not negotiable!}''. 

Given this background, it was no surprise that Paddy acquired a high
level of expertise in mathematics --- well ahead of what was taught in
his school, the Government Karamana High School, Trivandrum, where he
did his schooling in the vernacular, Malayalam medium --- and
developed a strong interest in geometry.
Other than his mathematical abilities, he was not a child prodigy of
any kind; while he was within the top three students in his class all
along, he was not even a consistent class topper in his school.
(His major problem was the Hindi language, which was compulsory and he
never quite mastered.)
Another passion during his school days was chess, which  he learnt
from Neelakanta Sarma.
Unfortunately, even state level competitive chess needed devoting time
which he could ill afford, and, at some stage, he decided to pursue
academics rather than chess.
However, over the last ten years he spent a
lot of time exploring chess again with the help of many computer chess
systems, and had an extremely high Elo rating against programs like
Fritz and Leela. 

After ten years of schooling (1963-72), he joined the Government Arts
College in Trivandrum for two years of the Pre-degree (1973-74) as it
was called then.
Three major events occurred during this period, which were
instrumental in determining the course of his future.  

First, he came across the \textit{Feynman Lectures in Physics}, and
found Physics to be more fascinating than pure mathematics, which was
the original career that he was contemplating.
``It appeared to me'', in Paddy's words, ``that theoretical physics
beautifully combines the best of objective science and the elegance of
pure mathematics.''
However, other than influencing his career decision, the Feynman
lectures did not figure in his physics education in a significant
manner; he learnt undergraduate physics from the 5 volumes of the
\textit{Berkeley Physics Course}, and later progressed to the 10
volumes of Landau and Lifshitz's \textit{Course of Theoretical
  Physics}.   

Second, he came across a wonderful organization --- and later became
an active member of it --- called the Trivandrum Science Society.
This was an organization entirely run by students from colleges in
Trivandrum, financed by membership fees and donations from
well-wishers.
Here, the members devoted themselves to the pursuit of science,
unshackled by curricula and examinations.
Paddy and a few others had an informal self-study group which
concentrated on theoretical physics, and in a span of about 3 years,
Padmanabhan managed to master the volumes of the \textit{Course of
  Theoretical Physics} by Landau and Lifshitz.
The process of self study and peer learning was empowering and this is
something that Paddy tried to inculcate in students and younger
colleagues. 
We have tributes from two colleagues from the society in this volume:
V. Parmeswaran Nair and Sarada G. Rajeev.

Third, Paddy took the National Science Talent Search (NSTS)
examination organized by the NCERT, Government of India, which was
probably one of the greatest stimuli that the government provided to
students who wanted to pursue pure science.
Success in this examination guaranteed a handsome scholarship for the
rest of one's scientific career, as long as one pursued pure science.
The money was very important to him as his  family's financial
position was never too good.
In addition, NSTS scholars participated in one-month summer camps at
leading institutes in the country, allowing them to interact with
researchers even while they were pursuing their college education.
The Trivandrum Science Society also used to run ``classes'' to prepare
students for the NSTS exam.
These classes were run essentially by senior students and sometimes by
one's own contemporaries.

After his pre-degree, Paddy joined the University College, Trivandrum
for his Bachelor's degree (B.Sc, 1974-77) in Physics.
His final year of pre-degree and the first two years of B.Sc were
again noteworthy in two respects.
First, this was the time when he worked through the Landau-Lifshitz
volumes and various other books in theoretical physics, spending
sometimes fourteen hours a day in this pursuit.
He also acquired a fascination for anything related to gravity and was
strongly influenced  by the epic book \textit{Gravitation} by Misner,
Thorne and Wheeler (Freeman and Co.).
He worked through the entire book, solving each and every problem
along the way. 
The second key event, which has nothing to do directly with his
academic pursuits but shaped his entire attitude towards life, was his
exposure to Upanishads, Zen philosophy, meditation techniques, etc.
He pursued these till the end of his life, being deeply
interested in problems of consciouness, meditation etc. as two well
known articles in his homepage (www.iucaa.in/\~{}paddy) illustrate.

The mastery of theoretical physics, especially General Relativity
(GR), helped him to publish his first technical paper in GR in 1977
when he was still a B.Sc student.
He started working seriously on several ideas in GR and quantum field
theory around this time.
He was a Gold Medalist for  topping the  B.Sc exam in
Kerala University, and joined for his Masters in Physics (M.Sc) in the
same University college.
Given the fact that he already knew all the standard stuff which was
taught in the M.Sc, he had sufficient spare time for his self driven
learning process and research work.
The interaction with the other members of the Science Society was very
helpful in these academic activities.  

Paddy was again a Gold Medalist for topping the
M.Sc. in Kerala university in 1979.
By now, he had caught  the attention of  several leading scientists
both in India and abroad.
The NSTS summer camps (at IIT, Kharagpur  and the Raman Research
Institute, Bangalore) as well as an Einstein's Centenary Symposium
(1979) at PRL, Ahmedabad which he managed to attend, helped
significantly in this regard.
Pursuing a Ph.D in the US or the UK (like some of his close friends in
the Trivandrum Science society did) would have been a logical course
to follow, but his family circumstances prevented him.
As a result, he decided to join what was then probably the best
research institute in the country (and an internationally acclaimed
one), viz., the Tata Institute of Fundamental Research (TIFR) for his
Ph.D.
He joined TIFR in August 1979 and became its tenured faculty member
(called Research Associate, which was the entry-level faculty position
in those days) in February 1980, while still working towards his
Ph.D.
His thesis work was in Quantum Cosmology (done under the supervision
of J.V. Narlikar) and he received his Ph.D degree  in 1983.
This work developed a particular formalism of quantum cosmology which
had the potential to  solve the cosmological singularity problem - an
idea which echoes in many of the currently fashionable  quantum
gravity models.
His thesis also contained  the notion of the wave function of the
universe, which was being developed independently by Hartle and
Hawking around the same time, from a different perspective.  

During his Ph.D, he met and fell in love with Vasanthi (who was also a
research scholar, one year junior to him, in TIFR).
They got married in March, 1983 when he had just completed his Ph.D
and Vasanthi was still pursuing hers.
She was working under the supervision of Ramnath Cowsik on the nature
and distribution of dark matter in the universe.
Paddy found this area fascinating and started collaborating with her
in this subject.
This also broadened his interest into several aspects of astrophysics
- an interest which has continued since then - and resulted in his
entering the area of cosmology, in which he later made his mark.
The book {\sl Structure Formation in Cosmology} was indeed a `labour of love'. 

In order to pursue his new research interests, Paddy decided to take
up a research associate position at the  Institute of Astronomy,
Cambridge for one year (1986-87).
He was strongly influenced during this time by Donald Lynden-Bell,
whose scholarship and breadth of scientific interests resonated well
with his own way of doing science.
He found the subject of the Statistical Mechanics of Gravitating
Systems particularly fascinating, and spent a fair amount of time
working on different aspects of it.
His own contributions to this area are highlighted in the first
single-authored review he wrote for \textit{Physics Reports} in 1990
and the later lecture notes of the prestigious \textit{Les Houches
  Schools} in 2002 and 2008.
His authority in  this subject is well recognized not only by the
astrophysics community, but also by the condensed matter community
interested in the statistical mechanics of long range systems. 

Soon after completing his thesis, Paddy started supervising two
students (T.R. Seshadri and T.P. Singh were the first two). The steady
flow of students continued --- in spite of him being rather selective
--- and he has so far supervised the thesis work of 16 students.
It is commendable that more than ten  of them hold faculty positions
in different institutes in India  and are guiding their own students.
At present his academic family tree has grand-students at different
stages of thesis work and at least a few have graduated!
A very significant fraction of young ($\leq 45$ years)
cosmologists working in various institutes/universities in India today
have been associated with Padmanabhan and mentored by him in the
Ph.D/post-doctoral stage of their career, in one way or another. 

Paddy started working on science outreach during his TIFR days and
continued to write on a variety on themes.
Paddy became a regular contributor to two science magazines of India
(the \textit{Science Today} and the \textit{Science Age}) which
existed at that time. He ran several regular columns in these
magazines, like \textit{Playthemes} (on recreational mathematics),
\textit{Let us think it over} (on everyday physics applications), and
\textit{Milestones in Science} (on the history of science). These
columns influenced generations of students including one of the
writers. 

His strong interest in the history of physics prompted him to present
the \textit{Story of Physics} as a comic strip serial in the
\textit{Science Age}.
This was extremely popular, and has been published as a book, translated
into several Indian regional languages and made available to the
school children at an affordable price.
This was made possible, in large part,  because Paddy did
\textit{not} take any royalties from this work.
Later, he wrote a 24 part serial in the science journal
\textit{Resonance}, called the \textit{Dawn of Science}, dealing with
the history of all sciences, from pre-history to the 17th
century. This has been published as a book now. 

 A popular science book which he wrote, \textit{After the first three
   minutes} [2000; Cambridge University Press (CUP)] taking off from
 where the book `The First Three Minutes' written by Steven Weinberg
 concluded, has also been very well received and was translated into
 Portuguese, Chinese and Polish. Paddy was always strongly committed
 to the responsibilities of scientists towards the society, and
 continued to be very active in public outreach programmes till his
 last days. 
Over time he also gave
numerous public lectures: we do not know of any instance when he
turned down an invitation irrespective of the scale of gathering.  

Around 1990, Paddy started working on his first single-authored book,
\textit{Structure Formation in the Universe} (CUP, 1993).
He was persuaded to write this book by Martin Rees, who introduced him
to  Rufus Neal, the CUP commissioning editor.
This book was extremely well received and made Paddy famous among the
astrophysics and cosmology community.
He took to writing books like a fish to water, and has published
eleven more --- with a few more in the offing!
Martin Rees had commented that he had triggered a run-away process
when he persuaded Paddy to write that first book.

It was a difficult thing to understand how he managed to write so many
high-quality books while keeping up with his research, and maintaining
an average of more than eight research publications per year.
Paddy explained it thus: ``Well, you need two things. First, you need
the discipline to work on it 2 hours each day -- in which you can
write 6 pages, if you have everything ready in your head.
So, a 600 page book will just take 100 days.
But, of course,  you won't have everything in your head, so it will
take about 5-10 times more time; so you can turn out one book every
2-3 years.
The second thing you need is Vasanthi.
She worked with me in all the books, taking care of much of the
typing, latexing and back-end processing!''
Every one of his books acknowledges Vasanthi's contribution. 
It was not unusual for friends to turn to Vasanthi while asking about
the progress in book projects!

In 1992 he shifted from TIFR to the Inter-University Centre for
Astronomy and Astrophysics (IUCAA).
During the early part of his career at IUCAA, he concentrated on
various aspects of structure formation in the universe.
During this time, he continued to work on his other interests centered
around quantum fields, gravity and approaches towards quantum
gravity.  

Coming from a theoretical physics background, Paddy's perspective on
astrophysics was rather different from that of many other people whose
initial training itself was in astrophysics.
In particular, he noticed that there was no comprehensive treatise
covering all of astrophysics, like, for example, the Landau-Lifshitz
course for theoretical physics.
He was lamenting about this to Jerry Ostriker, during his visit to
Princeton in 1996, when  Jerry asked him ``Why don’t \textit{you}
write it?''
During the next few weeks, Jerry was very supportive and helpful in
making concrete the structure of a 3-volume \textit{Course of
  Theoretical Astrophysics} which Paddy went on to author.
These three volumes were published by CUP during 2000-2002 and
reviewers have called them magnificent achievements.
Through these, Paddy became well known to a very large community of
astrophysicists and students  and his breadth of scholarship was
recognized all around.
In conjunction with  his  later books,\textit{Gravitation} (CUP, 2010)
and  \textit{Quantum Field Theory} (Springer, 2016)  he covered almost
all the frontiers of theoretical physics and astrophysics. 
\begin{figure*}
  \includegraphics[width=5truein]{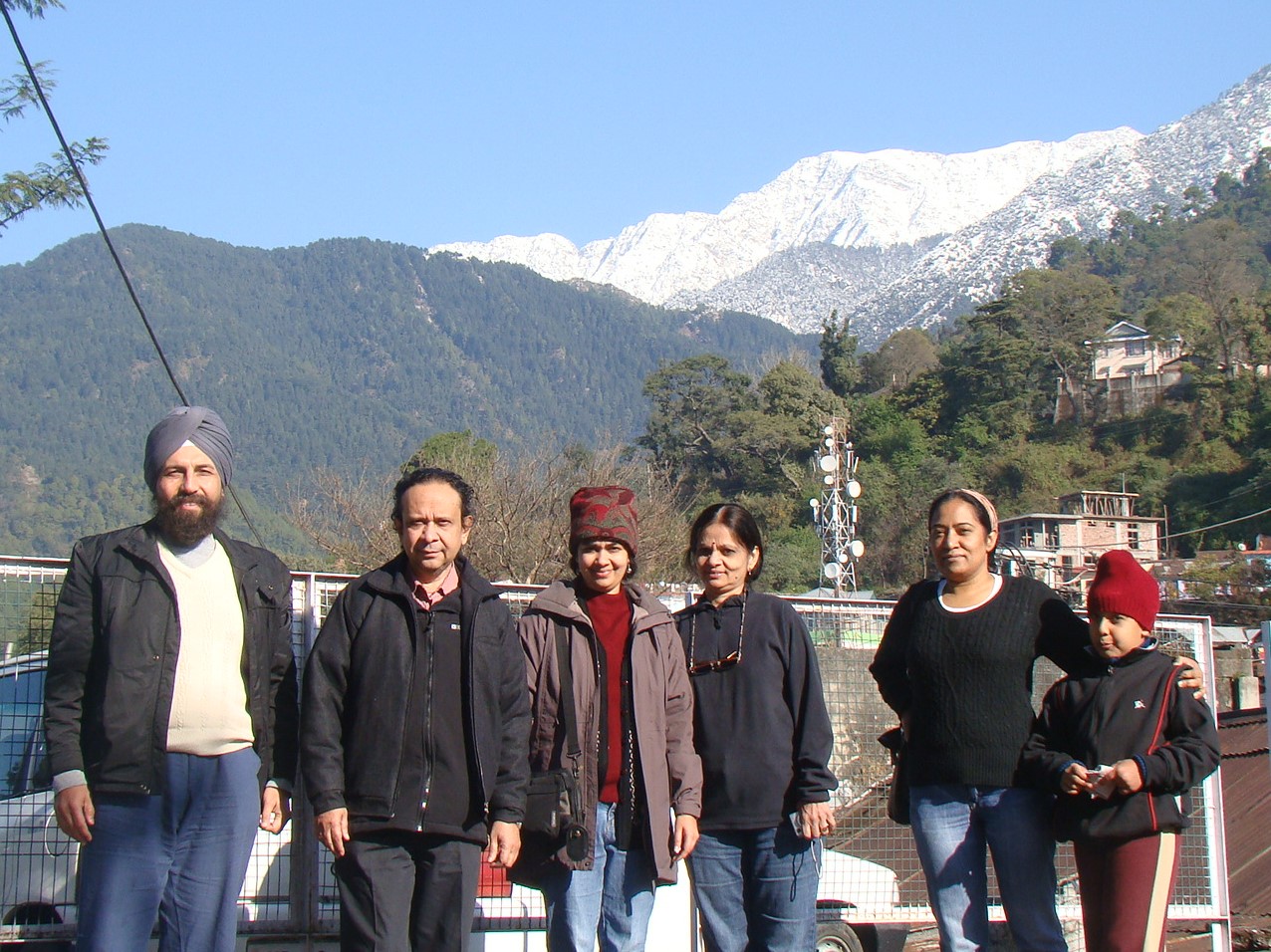}
  \caption*{Paddy, Vasanthi and Hamsa with Jasjeet, Harvinder and Janak
    in Dharamsala.  This picture was taken in December 2014.}
\end{figure*}

One reason he could write so many books was his passion for innovative
teaching. Paddy strongly believed that research and teaching should go
hand-in-hand.
He considered teaching to be an extension of the self learning process.
We recall an instance when summer students at IUCAA asked Paddy to
request one of his students to teach them general relativity.
He responded by volunteering to give lectures: what followed was a set
of six lectures given in an informal setting between 9pm to midnight. 
He has taught virtually every aspect of theoretical physics and
astrophysics at the graduate schools in TIFR and IUCAA, in addition to
occasional courses at the Pune University and IISER, Pune.
Most of these courses are available for the world to learn from.
He was a fantastic teacher and his lectures on even routine topics
were punctuated by  creative and original approaches --- something
which echoed in his books, which extended  his classroom teaching in
many ways. 

The course he was working on  prior to his sudden demise, was a course
for senior undergraduates which intended to  teach them \textit{all
  of} theoretical physics in about 150 lectures. This course was
intended to appear as a four volume book.  

While completely at home with any aspect of theoretical physics,
Paddy's real passion was for quantum aspects of gravity and finding
the interface between the two frameworks.
He never let go of this, having tasted blood in the early years of his
career.
From the turn of the millenium, he spent more time in this area, with
his astrophysical and cosmological interests taking a back seat.  

The two most important conceptual advances in theoretical physics,
made during the twentieth century, were General Relativity and Quantum
Theory.
However, all attempts to put together the principles of these two
disciplines have repeatedly failed, often after a lot of hope and hype
which accompanied  each attempt.
Paddy's work over the last two decades, suggests that this is because
we have misunderstood the nature of space-time structure, and are
applying the principles of quantum theory to the wrong physical
entity.
The approach followed by Paddy and others has been to show that
gravitation can emerge as an effective phenomenon rather than being a
fundamental interaction. 

Another apect of Paddy's career — which  set him apart from many other
scientists in his peer group — was his willingness and capability to
provide scientific leadership in various ways.
He served in several key committees and took a leading role in the
development of astronomy in India.
Here are just a few examples: (a) The Department of Science and
Technology appointed him as the Convener of the Advisory Group
(2008-10) to facilitate India’s entry into one of the international
collaborations building the next generation Giant Segmented Mirror
Telescopes. He played a key role in taking this initiative and
developing a consensus in the Indian astronomy community in this task,
which has now led India into joining the TMT. (b) He served as the
Chairman (2006-09) of the Time Allocation Committee of the Giant
Meterwave Radio Telescope (GMRT).  He introduced many innovative
aspects into its working and been instrumental in streamlining several
aspects of the GMRT time allocation. (c) He was the Chairman (2008-11)
of the Indian National Science Academy’s National Committee which
interfaces with the activities of the International Astronomical
Union. In addition to advising the Government on policy issues, this
also required him to coordinate the International Year of Astronomy
2009 activities in the country.
In the international arena, he was the President of the Commission 47
on Cosmology of the International Astronomical Union (2009 - 2012),
and the Chairman of the Commission 19 (Astrophysics) of the
International Union of Pure and Applied Physics (2011 - 2014).

Paddy received numerous awards and distinctions in
India and abroad for his contributions.
He was an elected Fellow of all the three Science Academies of India
as well as of The World Academy of Sciences. The national and
international awards received by him include the Padma Shri (2007),
the J.C.Bose Fellowship (2008-), the Inaugural Infosys Prize in
Physical Sciences (2009), The World Academy of Sciences Prize in
Physics (2011), the Millennium Medal (2000), the Shanti Swarup
Bhatnagar Award (1996), the INSA Vainu-Bappu Medal (2007), the
Al-Khwarizmi International Award (2002), the Sackler Distinguished
Astronomer of the  Institute of Astronomy, Cambridge (2002), the Homi
Bhabha Fellowship (2003), the G.D.Birla Award for Scientific Research
(2003), the Miegunah Award of the Melbourne University (2004), the
Goyal Prize in Physical Sciences (2012-13), the Birla Science Prize
(1991) and the INSA Young Scientist Award (1984).
His research work has won  prizes from the Gravity Research
Foundation, USA seven times, including the First Prize in the Gravity
Essay Contest in  2008.
He has had a nearly unbroken stint of prizes from the First to
Honourable mentions over two decades in the Gravity Research
Foundation competition. 

Paddy always felt that theoretical physicists were a fortunate lot.
In the preface to  \textit{Sleeping Beauties in Theoretical Physics}
(Springer, 2015), he made his view quite clear: ``Theoretical physics
\textit{is}  fun.
Most of us indulge in it for the same reason a painter paints or a
dancer dances --- the \textit{process} itself is so enjoyable!
Occasionally, there are additional benefits like fame and glory and
even practical uses; but most good theoretical physicists will agree
that these are not the primary reasons why they are doing it.
The fun in figuring out the solutions to Nature's brain teasers is a
reward in itself.'' 
These are, perhaps, the best words to remember Paddy by.

\smallskip

 In the following pages, Paddy is fondly remembered by his friends from college days in Trivandrum, and by  his graduate students and post-doctoral fellows from his days in TIFR and IUCAA.

\section{The Trivandrum Period 1957-1979}
\smallskip
\centerline{\bf {V. Parameswaran Nair} ({\it CUNY, New York})}
\medskip
\noindent Shock, sadness and a sense of tragedy were my reactions when I heard about the untimely passing of Padmanabhan. He was a great physicist, whose insights and contributions will find a place in a complete theory of quantum gravity, and his life will rightfully be celebrated in the years to come. The work of his students and collaborators as well as the frequent citations to his publications are a great testimony to the impact and significance of his research.  But to me personally, his passing was also the loss of a friendship which, in many ways, defined us both. Padmanabhan and I were college-mates for over five years, but even more than that, we had a close affinity of similar interests and similar values, a meeting of minds one could say. 
Along with Rajeev (S.G. Rajeev, now at the University of Rochester) and others, we talked, argued, worked through many advanced books on physics, told each other about new things we had learnt in physics. Yes, there was some sense of competition which drove us all, but ultimately, it was a conversation that lasted many years, sustained by curiosity. There was an intellectual exuberance and a sense of resonance when we talked. I think we were 
both lucky in the special friendship we had. I certainly felt so, I believe Padmanabhan would agree. 

Many memories surface when I think of those days, but a particularly vivid one is about a symposium organized by Professor Philip when Padmanabhan and I were undergraduate students at the University College, Trivandrum. During the lecture by one of the scientists from another university, Padmanabhan and I said some words, {\it sotto voce} of course, to each other concerning some point being presented. The speaker was annoyed and made it clear to us, and to everyone. We felt rather put down, so we started raising issues more openly, suggesting how one could use Schwinger’s action principle to resolve a point being made and so on, each of us taking turns with questions and comments. At the end of the symposium, all the invited speakers were eager to be introduced to us and left with a good impression of University College, I believe.

We were also very much involved with the Science Society in Trivandrum, which was an organization of students started a few years earlier by Arun Kumar (now CEO, KPMG, India) 
and his friends. We organized seminars by scientists we managed to invite, whenever we could, but most importantly gathered fellow students and talked to them, and talked with them, often more than once a week. It is impossible to keep to oneself the joy of the revelatory moment in learning or discovering something, we had to talk and share. Both Padmanabhan and I gave countless lectures on many different topics. Looking back, it certainly helped us in our own maturation as physicists; hopefully it helped our listeners to some extent as well. 

Padmanabhan’s interests were already slanting towards gravity even in those days, while I was more into field theory and particle physics. So naturally, our professional and personal lives diverged somewhat after I came to the U.S. and we did not manage to maintain regular contacts, more my fault than Padmanabhan’s. But friendships, such as what we had, which deeply influence our identity, do last a lifetime, no matter the distances. So I say a somber goodbye to a friend even as we celebrate his legacy.

\newpage

\centerline{{\bf Sarada  G. Rajeev} ({\it University of Rochester}) }
\medskip
\noindent {\it Memories of an old friend}

\noindent I met Padmanabhan (he was not called “Paddy” until he left Trivandrum) when he was a pre-degree student at the Govt. Arts College in Trivandrum and I was in tenth standard. It would be considered his last year of High School today, but this was before the plus two system was introduced. I met him at a meeting of the Science Society of Trivandrum, along with its chief, V. Parameswaran Nair. This was a student-run organization founded a few years before by Arun Kumar (now the CEO of KPMG, India).

Meeting them transformed my life. I had by then learned some physics and mathematics on my own, mainly because I wanted to find out how airplanes worked. Through them I heard about the Feynman Lectures in Physics and realized that physics explained not only airplanes, but every other amazing thing about the physical world.

Padmanabhan  and Parameswaran both totally amazed me  in that first meeting. Not just the depth and breadth of their knowledge, but their value system resonated with me. Everything had to be examined from the bottom up, taken apart and put back together in your own way. Trying to keep up with their conversation (what the heck is a “Hamiltonian”?) spurred me on to learn fast.

There were many other sides to Padmanabhan. Although not well off, he was from a far more  cultured background than me. It was a revelation to me that people could value knowledge for its own sake and not for a job or prestige it would bring them. He would tell me about the philosophical message underlying the Ramayana and Mahabharatha. That the commentaries of Shankara were in the form of dialogues with an imaginary opponent. That Kerala once had a great tradition in Mathematical Astronomy. Each of these insights have guided my life in the years since, although they were no more than casual comments for him.

We also shared an  interest in British detective fiction. The day I finally beat him at chess (after many, many losses) is a clear memory.

Those were years of intense competition and curiosity. I remember looking through the book on Classical Mechanics by Goldstein at the University Library. Someone had recently put a check mark next to various problems. It could not have been Parameswaran, as he had already learned it some years ago. So it had to be Padmanabhan, marking his own progress. I made it a point to solve not only the problems he had solved, but also the remaining ones. Among them was rigid body motion, a favorite topic to this day.

Our tastes in physics were similar, but not identical. He was more impressed than me by the   book by Misner, Thorne and Wheeler. I preferred the austere discussion of General Relativity in Landau Lifshitz (Classical Theory of Fields). In Padmanabhan's own book on Gravity, I see traces of some of our early arguments.

We left Trivandrum the same year: TIFR for him and Syracuse University for me. Our contacts since then have been sporadic. But not a week goes by that I am not reminded of some conversation from those early years. For example, last week, a colleague asked me a tricky question about Faraday's law. I knew the answer because Padmanabhan and I had argued about this back then.

Padmanabhan's legacy is not only his impressive research output and the excellent books. It is also the many people whom he inspired and trained to do physics. It seems so unfair that he will not be around another twenty years to create yet another generation of scientists. The only consolation is that  his students will continue the tradition. Indian physics will forever have the imprint of Padmanabhan in its DNA.

\section{Tata Institute of Fundamental Research, Mumbai 1979-1991}
\smallskip
\centerline{{\bf Urjit Yajnik} ({\it Indian Institute of Technology Bombay})}
\medskip
\noindent {\it Paddy in memoriam}

\noindent Paddy was a brilliant physicist, and a friend philosopher and mentor. His passing away has left a big gap in the firmament of astrophysics, with a special loss for India. 

I first met Paddy as far back as 1979 when he was an early PhD student at TIFR since Prof. J V Narlikar had kindly agreed to guide my MSc project at IIT Bombay and also, my senior Kandaswami had joined that group as a PhD student. I then went to the University of Texas at Austin for PhD, with my sights set on Quantum Gravity though eventually I worked in cosmology and grand unification. Just around the time I had defended my thesis, in 1986, my advisor Prof. E C G Sudarshan invited Paddy to visit our group. I became his default host and since as a graduate student I had been cooking, he relished my barely functional culinary skills. I was working on the proof of a missing link related to topological defects and phase transitions in gauge theories considered in my thesis and Paddy was quick to pick up the issues and helped with finishing the problem. This resulted in my only paper coauthored with him. 

He joked as he was leaving that my pasta skills were an added qualification for a postdoc at TIFR. I joined the group in January 1987 and Seshadri and T P Singh became close friends. My work concerned QFT in curved spacetime applied to inflation and my calculations grew in size and the results they yielded were unusual. Paddy was always encouraging and never at a loss and a big help in elucidating the subtleties. It was difficult to convince the referees but the paper finally got published in Phys Lett B. Even as recently as 2020 Paddy had said let us revisit the calculation.
\begin{figure*}[!h]
\centering
\includegraphics[width=12.00cm]{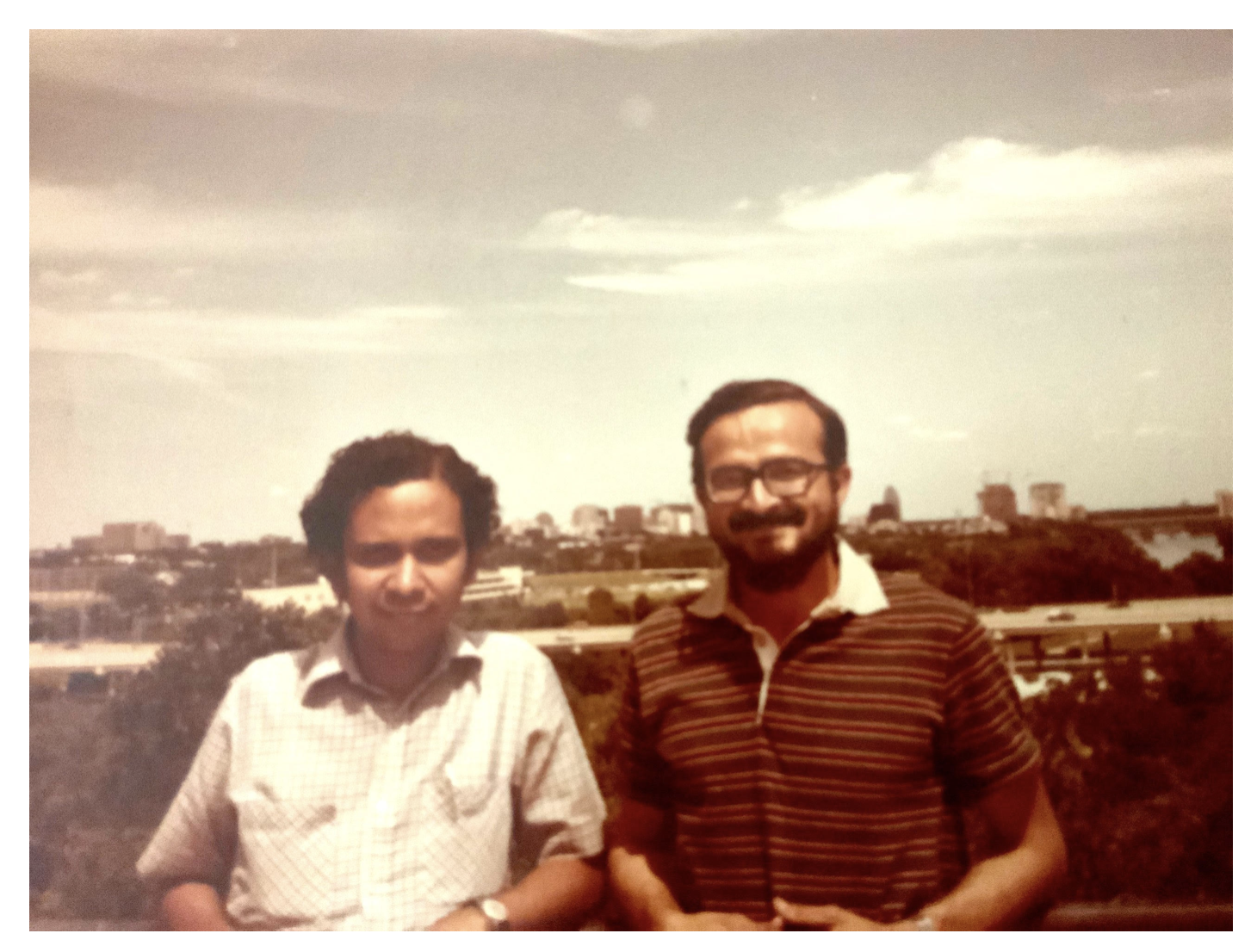}
\caption*{Paddy and Urjit in Texas, 1986.} 
\end{figure*}
Quantum Gravity was close to Paddy’s heart and my having worked with Weinberg, briefly with Wheeler and having taken courses from Bryce and Cecile de Witt and from Claudio Bunster (then Teitelboim) provided a strong link over which we bonded.  In 1989 Paddy lectured on Quantum Field Theory at a winter school in Pune and he invited me to join as a “tutor”, but giving me a blank cheque so that it was possible to cover Weinberg’s perspective on QFT, relativistic wave equations and conditions for the causality of the S-matrix, at that school.  In the same year, at a meeting on Gravity he invited me to join a panel discussion with himself and Ashtekar, about issues in quantum gravity.  

After I joined IIT Bombay, during all the visits to IUCAA it was a pleasure to meet up with Paddy and exchange many jokes, viewpoints and of course physics insights. He was always working on something uncanny and pushing ahead developing ever new ab initio perspectives on semiclassical gravity.  After the precision data on Cosmic Microwave Background by the WMAP experiment in 2003 we had the first ever confirmation of how the quantum ripples of the Big Bang, specifically an inflationary phase of the early Universe could be the cause of our origins. Paddy held a special focused meeting under the aegis of the Indian Academy of Sciences with about eight participants towards understanding all the aspects of these new developments. At this meeting, my preliminary slide said, “inflation by definition was not supposed to leave behind any signature”. This amused him a lot and he said he would borrow this pronouncement, and he would bring it up in our conversations for quite a few years. 
\begin{figure*}[!h]
\centering
\includegraphics[width=12.00cm]{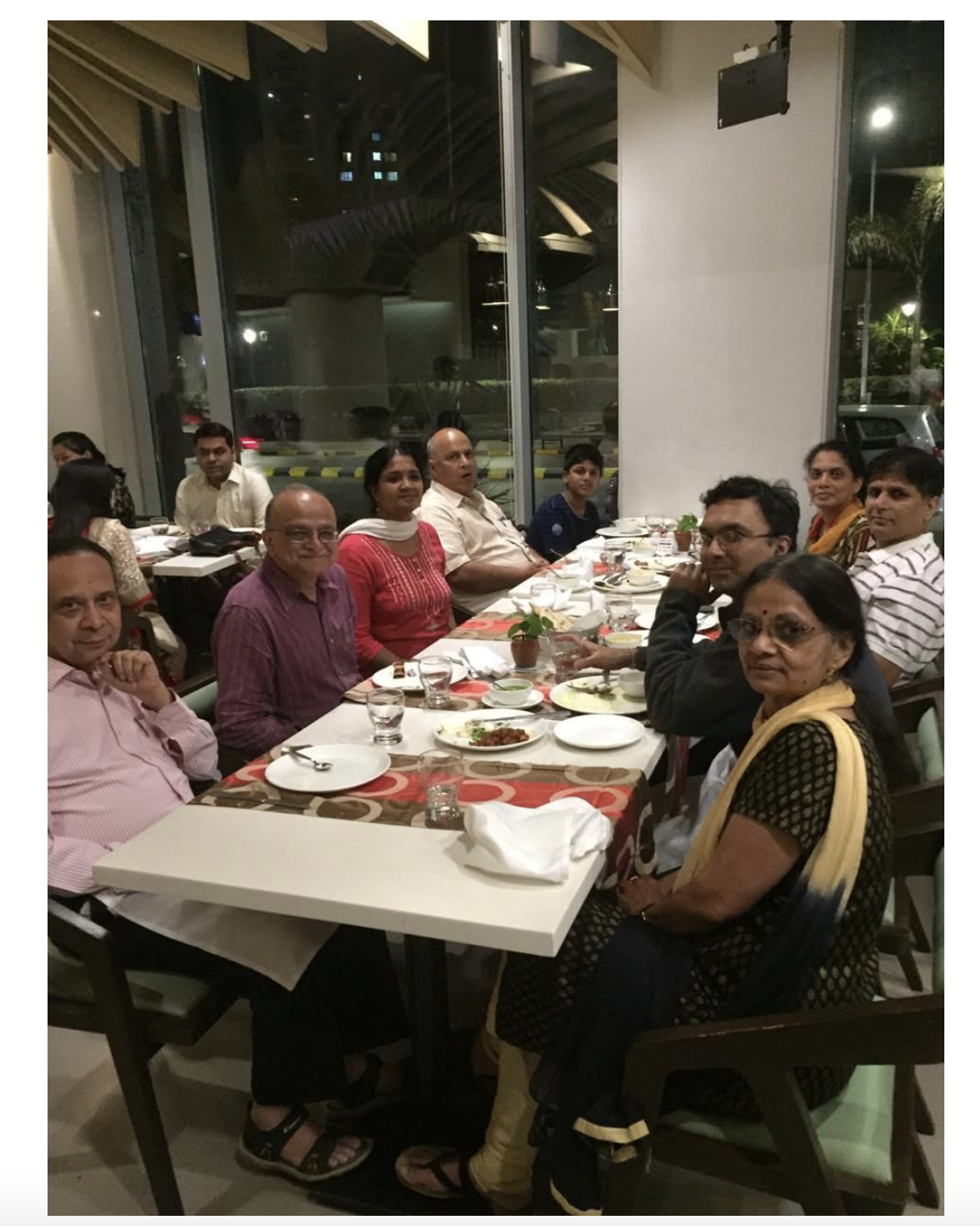}
\caption*{Paddy and Vasanthi visiting IIT Bombay, 2019.} 
\end{figure*}
A very nice conference was organised to mark Paddy’s 60th birthday and it was a pleasure to make a presentation, linking up some work later done by my group that Paddy had helped with back during the Texas visit of 1986. My contribution on this occasion, titled “Stability longevity and all that : false vacua and topological defects” covers a few other vignettes from his Texas visit of 1986.

Another recent occasion to meet was when he accepted to deliver a lecture in the Diamond Jubilee Distinguished Lecture Series in our Department, and both he and Vasanthi spent some time on our campus.  His colloquium as usual was a grand success and kindled the interests of a large number of students. Our last occasion was Karthik’s defence at IUCAA in February 2020 when I also got a sumptuous dinner at home and the gift of his book on Quantum Field Theory with a fresh perspective and some rather intriguing derivations, written in the format of a self study book for a student. I studied the book during the lockdown and was hoping to catch up with him on many of the topics when movement gets restored. Unfortunately we have lost him before this could happen.

Some consolation to us is the substantial body of work he leaves behind, including many expository articles and books as well as popular science  presented with light hearted stories. This will continue to provide us clues to future studies and will inspire many young students in the coming generations. 

\bigskip
\bigskip

\centerline{{\bf T. R. Seshadri} ({\it University of Delhi})}
\medskip
\noindent {\it Paddy - My Friend}

\noindent Paddy was a genius - not just in Physics but in a whole spectrum of knowledge and activity. He had his first publication as a B.Sc. student. He joined TIFR in 1979 for PhD and became a faculty member within a year. I came in contact with Paddy for the first time in 1983 when I joined TIFR for my Ph.D. He was just 4 years senior to me. We became good friends and used to talk about Physics, History, movies and many other things. Often we would chat over coffee and lunch in the West Canteen. During one such conversation, he asked me what I planned to work on. I was interested in Cosmology, and it occurred to us that since he was already in the faculty, I could become his student. Those days TIFR students were generally registered in Bombay University and there was a formal procedure for a TIFR faculty to be recognized by the former to supervise PhD students. Paddy had not applied for it. So Jayant Narlikar, who was the head of the group suggested that I register with him and then shift to Paddy once the formalities of his recognition as guide got completed. . This was the way I had the fortune to become the first PhD student of Paddy.

Gradually I became very close to his family. Often we will be chatting till late evening and I will have dinner at his place. As before conversation would span anything from movies to serious Physics discussions. In fact a few of my papers during PhD evolved mainly around such Physics discussions. Often, especially on Sundays, we (Paddy, Vasanthi, Ramakrishnan, Rajaram and I) used to take the 7 PM TIFR bus, have dinner, see a 9 PM to 12 midnight movie in a cinema hall and return.

Paddy, Vasanthi, and Paddy's parents would treat me as a part of their family. In fact if some dish was prepared for lunch at their place, Paddy's parents would keep it for me and send a message through Paddy and Vasanthi that I could come and have it during lunch. Later, when Hamsa was still small, she would insist on playing with me whenever I went to their home in IUCAA.

Paddy was interested in anything that involved intellectual pursuit. He tried to get me interested in these. Some I could manage. Some I couldn't. Chess was an example of the latter. In fact he even got me a chess board and a book on chess but I think I didn't have the intellectual capacity to play chess. One subject he got me interested in is History. Among the many conversations ever etched in my mind is the one about history during a dinner in the East canteen. He asked me if I liked history. I gave what I thought was a careful answer. I said I didn't like History the way it was taught in school. Paddy immediately asked if I liked the way Physics was taught in school. I said "No". But I already realized what he was heading too. His immediate question was how I was then interested in Physics. This was an important point that I had not thought of. In course of time we had started discussing history and I have a strong interest in it even now. The credit for this completely goes to Paddy and the logical way he analyzed and historical events that we discussed.

His approach to Physics influenced anyone who came in contact with him. For whatever I learnt in Physics, Paddy has had a very significant role. He not only taught us to do well in physics, but importantly taught us how to enjoy physics. It is mainly due to him that many of us who have been his students could work in diverse areas of Physics. Apart from research he also exposed us to the importance of pedagogy, something which has helped me a lot during all my years of teaching. One of the examples of this was getting us addicted to Landau and Lifshitz.

When I joined TIFR, I was an introvert. It was one of the greatest contributions of Paddy in my life that he gave me the confidence to stand on my own and swim against the flow without hesitation when needed.

A banyan tree drops roots, which in turn become full fledged trees. The nurturing by the parent tree ensures that the cluster lives on long after the parent tree is gone. Paddy is not physically with us anymore, but he has trained the members of the Gharana so well that we can continue that tradition, That would be the greatest tribute we can pay to this genius, this friend.

\bigskip
\bigskip
\newpage

\centerline{{\bf Tejinder P. Singh} ({\it Tata Institute of Fundamental Research, Mumbai})}
\medskip
\noindent {\it This is a short letter written to Paddy on 25th September, 2021. I imagine that he must have read it.} 

\noindent Dear Paddy,

\noindent On the 17th of September, around 11 am, Professor Seshadri phoned in to say that you have left us. Since then, these nine days have been days of grief, disbelief, and memories. When I go to the office, I walk past the office room you used to occupy in TIFR. When I look out from the window of my house, I see the house you lived in during your TIFR days. Your home was Gurukulam, and you and Vasanthi and your parents treated us students as your family members. Happy times were spent playing with Hamsa when she was just an year old. Our deepest condolences to Vasanthi and Hamsa, who are themselves dear family friends to my wife Jyoti and me, since our TIFR days together.

Paddy, on the physics front, when I examine my current research on quantum gravity, I see your imprint all over. During 1984-1989, you were my Ph. D. Supervisor. Along with Seshadri, and Urjit Yajnik, who was your first post-doc, we had a small but intense and close-knit group. Since you were only five years older than me, you were also a close friend. We spent an endless number of hours together, doing research, discussing a large variety of problems in theoretical physics, and discussing matters of life in general. The following are the three most important physics lessons that you taught me.  One, physics should be done for the joy of doing physics, not for building a career. Two, on any given physics problem, one should develop one’s own original line of thinking. Three, one should become a broad-gauge physicist, and not just continue working on a narrow topic.

After you left for IUCAA, our interaction reduced, but you had already done the crucial molding that shapes a student into a scientist. At IUCAA, you did your great work on gravity as thermodynamics, which will decidedly leave its mark on the final theory of quantum gravity, as and when the final theory comes. 

Paddy, you leave behind a gharana of some thirty former Ph. D. students and post-docs, who have had their own students. Also, Vasanthi and Hamsa are  a part of this academic gharana. It is our solemn promise that we will take your scientific legacy forward and strive to achieve the high and exacting standards you have set for us. You may have left us, but we will not let go of you. Your books, research articles, and video recordings of your lecture courses are with us. If we have a question for you, we will look up the monumental works you have left behind. You have inspired thousands of college students, who have in the last few days expressed their condolences, and respect and admiration for you. It is our wish and hope that we will be able to set up a `Thanu Padmanabhan Centre for Gravitational Physics’. At this centre, your gharana will work to train these upcoming youngsters, and in so doing, everlastingly keep you in our memories. That is a promise Paddy, we will never forget you. Your life was short, but it was extraordinary. Trivandrum, Mumbai, Pune, and now a bright star in the sky. We feel your light Paddy. Keep shining in on us, my teacher, my friend.

\section{Inter-University Centre for Astronomy and Astrophysics, Pune 1992-2021} 

\centerline{{\bf L. Sriramkumar} ({\it Indian Institute of Technology Madras})}
\medskip
\noindent {\it Memories of Paddy}

\noindent

\noindent Professor Thanu Padmanabhan, affectionately known as Paddy to many of us,
was an intellectual giant and each of his students have their own versions 
of experiences with him, much like the blind men and the elephant. 
The sudden demise of Paddy has been a devastating loss to us. 
He is irreplaceable.
Paddy's limitless energy and enthusiasm were admirable. 
As I think of Paddy, I am flooded with many memories and I will share three 
distinct ones.
\begin{figure*}[!h]
\centering
\includegraphics[width=12.00cm]{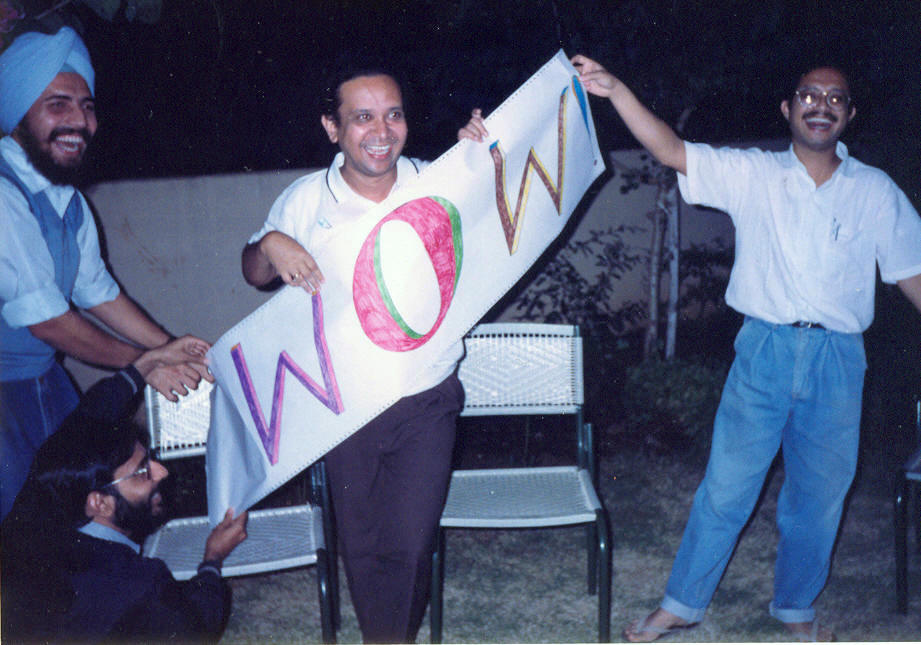}
\caption*{Paddy (at the centre) after he had won the Birla Science Prize in 1992.} 
\end{figure*}

I should consider myself fortunate and privileged to have known Paddy and 
have had access to him. 
The access was simultaneously beneficial and challenging. 
The access was beneficial for the reason that it helped me to constantly 
raise my own levels and challenging because of the fact that Paddy was 
difficult to keep pace with.
I will highlight a challenge and a benefit. 
When I had joined Paddy to pursue my thesis work, he had given me a list of 
problems to work on. 
I could complete a set of them, but the remaining ones were difficult. 
Paddy had already solved the problems I had solved and I had asked Paddy to
give me fresh problems that I can solve. 
He simply said ``If I can solve a problem, why should I give it to you!" 
I have always considered Paddy's stand as fair.  
Let me now point out a benefit of working with Paddy. 
I was writing my first paper and, in the context of the Unruh effect, I was
examining the response of detectors which are switched on for a finite time. 
I had encountered divergences in certain situations and I was struggling to 
understand their origin. 
A brief discussion with Paddy immediately helped me resolve the issue. 
He quickly recognized the source of the divergences and provided a proof
within a matter of minutes. 

It was Paddy who introduced us to the ten volume Course on Theoretical Physics 
by Landau and Lifshitz. 
For his students, it set the standard of Paddy's expectations. 
I also believe it is the Landau and Lifshitz series that inspired Paddy 
to write  the three volume set of books on Theoretical Astrophysics.
Paddy also read widely. 
He was the one who had introduced me to gems such as Gore Vidal's Creation.

Moreover, Paddy cared deeply for students and the teaching of physics. 
I believe that it is one of the reasons he was keen on writing textbooks. 
About a week prior to Paddy's passing away, we had the general body meeting 
of the Indian Association for General Relativity and Gravitation (IAGRG). 
One of the items on the agenda was making lectures on General Relativity 
easily available to a wider set of students. 
Paddy had tried to call me before the meeting, but I could not get to speak 
to him. 
I noticed his call after the meeting and had spoken to him. 
He said that IAGRG should in no way limit the courses and should focus on 
facilitating access. 
He had said ``let a thousand flowers bloom", which is what IAGRG had converged 
on in the meeting as well.

Paddy made impatience a virtue. 
As he had written in the preface to his three volume series of textbooks on 
Theoretical Astrophysics, he wished to cover as much area as possible within
a short span of time, much in the manner of Gengis Khan. 
Unfortunately for us, he seems to have exhibited the same impatience in his 
final moments too.
Paddy's passing away leaves behind a huge void in our lives. 
It will take us quite a while to come to terms with it. 
He will remain in our memories forever.

\bigskip

\bigskip

\newpage

\centerline{{\bf Tirthankar Roy Choudhury} ({\it National Centre for Radio Astrophysics, TIFR, Pune})}

\medskip

\noindent I first heard about T. Padmanabhan (or Paddy, as we called him) during my days in the university, perhaps when I was doing the M.Sc. Some of us got interested in Cosmology and we were told by one of our teachers that the quickest way to learn it is to solve the problems given in the “Cosmology and Astrophysics through Problems” by Paddy. Needless to say, we could not make much progress, however, we were all extremely impressed by the way the problems were organized in the book.

I joined IUCAA as a Graduate Student in 1998. We were taught by Paddy in the first year of the Graduate School. During the interactions, I had made up my mind that I would like to work with him for my Ph.D., except I was not sure if he would agree to take me. When he agreed, it was a dream come true.

I was truly fortunate to work on several different topics during my Ph.D. The primary was modelling neutral hydrogen in the Universe as probed by quasar absorption spectra (in collaboration with R. Srianand). In addition, Paddy gave me problems related to dark energy and the accelerating Universe (as the field became very popular with the new data from Supernova Type-Ia surveys). We also wrote a couple of papers on semi-classical gravity. Even now, I find it difficult to believe that the same person could not only work but also supervise Ph.D. students on various diverse topics so seamlessly.

In fact, one of the semi-numerical codes we (i.e., Paddy, R. Srianand and I) developed was to simulate the distribution of neutral hydrogen in the intergalactic medium. Based on some previous works, we found a neat way to make a simple model which could bypass many of the complications a full numerical simulation would face. Our formalism was computationally fast and hence could be useful to probe the unknown parameter space very efficiently while comparing with observations. The importance of the work can be understood from the fact that, even now (more than 20 years after it was developed), we keep on using it for many interesting calculations.

I completed my Ph.D. in 2003 and decided to continue on issues related to neutral hydrogen and galaxy formation. Paddy, around the same time, embarked on his ambitious project of studying Emergent Gravity and so our fields of research diverged. I still wish Paddy to have continued on his astrophysics and cosmology research, I am sure he would have made some more fundamental contributions to the field. Personally, I perhaps would have interacted and collaborated with him more than I actually did. Thankfully, I did manage to discuss science with him later (around 2012) when Hamsa, his daughter, started collaborating with me.

Much of the way I do research today is inspired by Paddy. I try to follow his mantras of hard-work, discipline and simultaneous expertise in diverse areas. In many tricky situations in my academic and professional life, I was fortunate to be able to talk to Paddy and find a way to deal with. In fact, till recently I used to ask for suggestions from Paddy whenever I had to teach a course in the IUCAA-NCRA joint Graduate School. It is just too difficult to accept that the constant guidance would not be there any more.

\bigskip

\bigskip

\centerline{{\bf Sujoy Modak} ({\it Universidad de Colima, Mexico})}
\medskip
\noindent {\it Post-doctoral Fellow with Paddy, 2012-13}

\noindent Personally, it is deeply disturbing for me to switch from writing a research article in progress with Paddy, to a memorial article on his demise, just in a matter of few days. Paddy’s absence is a major setback for the global scientific community which has lost one of its distinguished flag bearers who has guided multiple generations of science aspirants to pursue a career in scientific research, education, and communication. His unmatched legacy will survive through his work and will continue to inspire generations to come.

In India, it is hard not to hear about  institutions like IUCAA and scientists like Paddy for students who aspire to be the next astrophysicist, cosmologist, or theoretical physicist in general. I was no exception. I knew about people in IUCAA with more detail, including Paddy, when I was a Masters student at the University of North Bengal (NBU), Darjeeling, which at that time had a departmental library named “IUCAA Reference Center”. During my PhD days at the S. N. Bose National Centre, working on semi-classical gravity, Paddy’s books and review articles were essential weaponry in the process of grooming myself into a theoretical physicist. I cannot remember any of my research articles during PhD which did not cite Paddy’s works. In the summer of 2010, into the third year of my PhD, I received an email from Paddy who seemed to be happy with the results of one of my papers, which demonstrated the validity of a thermodynamic identity for black holes using a different path. Obviously, Paddy had the same conclusion several years ago using another path. This was the first time when our paths had crossed to the extent that he took a notice of my work. That was one of the happy moments of my PhD days. Later I had the opportunity to attend his talks at S.N. Bose Centre and a few other places.

After finishing PhD in 2012, I came to IUCAA as a postdoc with Paddy. I clearly remember our first meeting - his momentary friendly gesture welcoming me at the IUCAA and then immediately going back into his work-desk. At that time, I decided to review a few areas of gravity related research and I told him I am not sure which direction I wish to pursue. Paddy gave me time to come up with some ideas and I did. Some of them were interesting to him and suggested me to pursue by myself. We also agreed to do a project together which included Suprit, his then PhD student. We concluded that work by the middle of 2013. While at IUCAA, I witnessed an utmost respect and awe directed towards Paddy, from the students and postdocs alike. Some of them often joked that it is not a good idea to speak to him for a longer period on a one-on-one basis because there is a danger of overwhelming yourself with new ideas of research and ending up working even more! Once, he asked me for my hard drive, which had  a fair amount of movie collection, and mentioned I should separate the “questionable contents” in a separate folder. Whether there were such contents on my hard drive is still an unsettled issue. 

By the middle of 2013, I got an extension at IUCAA for two more years, and, I had a couple of concrete possibilities to pursue another postdoc abroad. That was a very confusing time – on one hand I wouldn’t want to leave a postdoc with Paddy just after a year and on the other hand there was a norm that one must have a research experience abroad to get a position in India. I decided to take up the matter with Paddy and surprisingly he encouraged me to take up a position abroad. In September 2013, I left IUCAA and came to UNAM, Mexico, a country which I adopted for my future residence in 2016. Looking back now, I see how a suggestion from Paddy had shaped my life in a positive direction, just like it did for many others.

Irrespective to this, in my mind, I never accepted my early departure from IUCAA, and regretted my lost opportunity to learn more from Paddy. After many years, only recently, I started collaborating with him once again on a research project that we talked about a few years ago. We started conversation over telephone and exchanging notes (from my side) and ideas (from his side). Last email conversation with me was on September 6, 2021, where he wrote, in his style, “i will get back to you in due course” – it is heartbreaking for me that it did not happen, and I must wait forever for his next email!

\bigskip

\bigskip

\centerline{{\bf S. Shankarnarayanan} ({\it Indian Institute of Technology Bombay})}
\medskip
\noindent {\it A magnificent star transformed into a black hole}

As Charles Dickens would say, “This is the best of times, and this is the worst of times” for researchers, students, and teachers in gravity and the physics community at large! Astrophysics, Cosmology, and Gravity are in the renaissance! In the last few years especially, the research on Gravity has taken a new zeal after a range of discoveries. Unfortunately, we then hear  of the passing away of Paddy --- an original and thought-provoking scientist in gravity!  

Besides being a great thinker, Paddy was excellent at breaking a physics problem into small atomic units. He was persistent and meticulous in finishing each atomic unit before proceeding with the next atomic unit! I still remember when I just joined Paddy as his Ph.D. student, he decided to write the three-volume book on Theoretical Astrophysics and completed it in three years! I asked him naively, how can you finish three volumes in three years? He said, “Every day, I write ten pages. Out of the ten pages, only the contents of 3 pages will finally go to the book. If I do that for 150 days, I will have a 450 pages book.” 

Paddy’s life reminds me of the quote by Albert Einstein “Creativity is intelligence having fun.” For the past few decades, Paddy tirelessly continued his quest.  He seamlessly switched from research to lectures, writing books, mathematical puzzles, chess, music, and philosophy.  

Paddy was frank in giving his view or opinion on the research. Two recent stories come to my mind: When Paddy visited IIT Bombay in January 2020, one of my Ph.D. students presented his current work. He told the student and me, “This is a completely wrong approach to take! Do it properly.” We took Paddy’s comment seriously and completed the work. When we sent the preprint for his comment, he called me and said, “I like the table in the paper”. During the same visit, Paddy and I discussed his work on emergent gravity and my recent paper with a former postdoc. He told me, “Do not try for a fancy journal! Give all the calculations and make it a readable paper.” Paddy’s brutal honesty reminds me of Thomas Jefferson’s quote, “Honesty is the first chapter in the book of wisdom".

Paddy was an excellent mentor and always considered his friends and students as his extended family. He was always  willing to help them whenever they asked for his help professionally or personally. When my father was very critical and needed advice, the only person who came to my mind was Paddy. It was late at night; however, he patiently discussed with me the problem and the plan to follow, at length.

Paddy’s homepage has the catch-phrase from Star Trek “Beam me up, Scotty! There is no intelligent life out here”. Today this phrase has taken more meaning to me than before! Suddenly, a magnificent star transformed into a black hole. All of us will gravitate around this black hole without seeing any light. I will miss a great friend and a true physicist! 

\bigskip

\bigskip

\centerline{{\bf Dawood Kothawala} ({\it Indian Institute of Technology Madras})}
\medskip
\noindent {\it My compass in spacetime}

\noindent My PhD thesis acknowledgement for Paddy had the following quote by the eminent Russian mathematician V. I. Arnold (speaking of a book by Isaac Barrow): \textit{For modern mathematicians it is generally difficult to read their predecessors, who wrote: “Bob washed his hands" where they should simply have said “There is a $t_1 < 0$ such that the image $Bob(t_1)$ of the point $t_1$ under the natural mapping $t \to Bob(t)$ belongs to the set of people having dirty hands and a $t_2$ of the half-open interval $(t_1, 0]$ such that the image of the point $t_2$ under the same mapping belongs to the complement of the set concerned when the point $t_1$ is considered.} Paddy liked the quote, and those who have worked with him would understand why it aptly captures a glimpse of a scientist whose insight, intuition, and scientific breadth and depth let him see through the mathematical clutter and identify if there is a ``basement to the building" [Everything in quotes are Paddy's words].
\\
\\
Paddy no longer being around is a very personal loss for me. As the news and the shock sink in, the thought that lingers is that I have lost someone with whom I could talk physics, from basic electromagnetism and classical dynamics to advanced issues in general relativity and quantum gravity, in a manner which kept alive that basic sense of awe which pull most of us into research in the first place. We were talking about Lenz's law less than a week prior to September the 17th. We got there as a result of our discussion on some geometric aspects of active diffeomorphisms. This, in turn, was led on by a discussion about how one would apply the second law of thermodynamics in the context of his proposed definition of the {\it heat content} of spacetime! It is this immense breadth of topics, and a common thread often running through them, that made all those long phone calls with Paddy so stimulating. We shared a common enthusiasm for pedagogical expositions and hidden curiosities in elementary physics and math topics. Paddy reinforced in me the fun and importance of looking deep into ``well understood" results. His style infused a certain strength to pursue paths less fashionable in search of fresh insights. One finds an echo of this in the following remark from his homepage: `` \ldots \textit{though progress has become very difficult in the recent years since this area has become fashionable!}" 
\\
\\
For a long time now we have been discussing the relevance of some lesser known tools of differential geometry to characterise the small scale, mesoscopic, domain of spacetime. Few days before he passed on, he told me about a new seed of an idea he had, related to a recent mathematical result in a paper I had written with a student. It took Paddy just one phone conversation to understand a completely new mathematical tool. In the second one, he got back, in his characteristic style, with a highly innovative physical application of it to certain issues in quantum gravity! Paddy was very keen that we should explore it seriously and put the tool to some ``noble cause". This week, we would have been discussing it further \ldots
\\
\\
My journey with Paddy started with a graduate school project on the so called River metric, and that project report also happens to be one of the last things we discussed. His last email in this regard (to Sumanta and me) had the subject line ``no". That was 16th Sep, 5:28 pm. The sheer suddenness of Paddy's next move has the aura of the last, grand disappearing act of a great magician. 

\bigskip

\bigskip

\centerline{{\bf Sanved Kolekar} ({\it Indian Institute of Astrophysics, Bangalore})}
\medskip
\noindent I first heard about Paddy through one of my professors during my Masters in a reply to my question “I want to pursue my Phd in India in the field of gravity, who should I work under/with?”. “Paddy in IUCAA. He is the Einstein of India. Anything related to gravity, he is the person to discuss with” was the answer. In what followed, I consider myself fortunate to be guided by Paddy during my time at IUCAA and witness in person the remarkable brilliance of Paddy. I remember the first time I \st{met} saw Paddy in the corridor of IUCAA during my graduate school, I was too scared even to say hello, scared in case I don’t say something clever enough for him and ruin my first meeting impression. It was during one of the courses on statistical physics which Paddy taught in the graduate school that I found the courage to talk and interact with him and then things went smoothly from then onwards for my sake. After joining under him as his student I distinctly recall the feeling I had, after each meeting in his office; I used to emerge escatic having learnt something new and creative. His way of interpreting and handling seemingly complex ideas and turning them into manageable or easy ones by breaking them up to their basic core is something I strive to emulate in my work as well. He was no doubt one of the best teachers I have met so far. His clarity of thoughts, immense knowledge and right-on-target scientific intuition has always amazed me. Other than the stimulating physics discussions, I also thoroughly enjoyed talking about puzzles with Paddy, be them either mathematical or logical or both, during the occasional dinners we had together at his house or in restaurants in Pune. 

Paddy attended my full wedding ceremony in Mumbai along with Vasanthi and Hamsa by travelling to and from Pune in the same day. Even though the wedding was mine, not surprisingly, Paddy became the main centre of attraction for my guests when he was there. An extremely busy and productive scientist of his calibre taking an entire day off just for me was a big deal for me and indirectly showed his affection towards his students. At IUCAA, we had an open invitation for food at his home whenever we felt like. When a terrible personal tragedy struck in my life, Paddy was always there for me as a fatherly guardian, looking over and consoling as well as encouraging me in his own blunt no-nonsense logic style, over many phone calls we had and also through the important phone calls he made for me.

I have gained a lot from him not just academically but even on a personal level which he may not be aware of and I am truly indebted to him for that. Now, suddenly he has left us all, in a shock. Once again, I feel like being orphaned for the second time. Though, I am sure his values, ideas and teachings shall remain in each one of us like his books and like the atoms of spacetime pervading all space and time. 

\bigskip

\bigskip

\centerline{{\bf Sudipta Sarkar} ({\it Indian Institute of Technology Gandhinar})}
\medskip
\noindent  {\it Remembering Paddy}

I first heard about Paddy when I was a physics MSc student at the University of Calcutta. After long deliberation, I chose general relativity as an elective paper. I was fascinated by the physics of curved spacetime and decided to work on gravitational physics. Then, I was immediately told to follow the works by Prof. Thanu Padmanabhan, the leading expert on everything related to gravity.

I joined IUCAA in 2004 as a beginning graduate student, and the first interaction with Paddy happened during the `General Relativity' course. The course was famous for being tough, and we were all warned repeatedly about the difficulty level. Also, the first few classes were a bit scary. Our instructor was one of the world's foremost theoretical physicists, and he also can write with both hands! But, eventually, the mathematical derivations, the physical arguments, the clarity of the explanation, and above all, the artistic board-work led to a magical experience. It was immediately apparent that we are dealing with a person who belongs to an entirely different league. 

I cherish the experience of working with Paddy. His approach to the students was unique. He always treated us as his friend and colleague. The lengthy discussion hours in his office have taught me how to effectively think about a research problem and tackle challenging mathematical derivations. Paddy's mathematical skills were legendary, but he always preferred a physical approach based on physics intuition. Mathematics only came as a necessary tool.  In my opinion, his remarkable power of physical intuition puts him at the highest level.

My Ph.D. work with Paddy was mostly on horizon thermodynamics. This is the time Paddy started his new research program of emergent gravity.  I feel fortunate that I had interacted with him very closely during this period.  I am also proud that I have coauthored his initial papers on the thermodynamic approach to gravity.

It is probably unnecessary for me to mention Paddy's influence on the Indian scientific community. In fact, from the beginning of Paddy's research career, it was clear that he was a natural leader without any parallel. He has set a standard for scientific research in India. Above all, he has taught us the importance of uncompromising academic honesty and integrity.

\begin{figure}[H]
\centering
\includegraphics[width=0.3\textwidth]{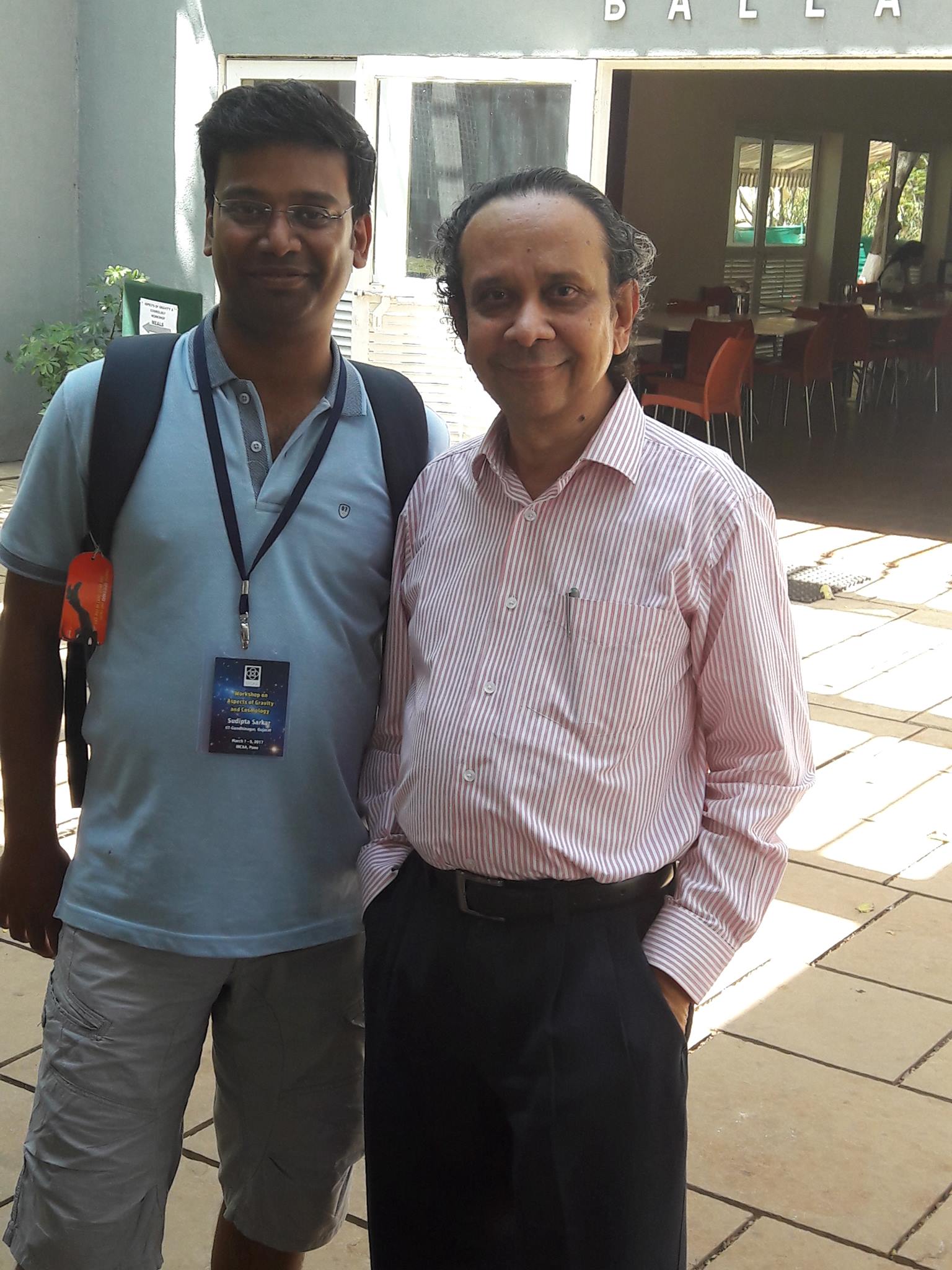}
\caption* {Picture taken during Paddy's 60th birthday celebration at IUCAA.}
\end{figure}

There are so many beautiful memories with Paddy. So much I have learned from interactions and arguments with him, including the art of enjoying south Indian filter coffee. In later years, whenever we met, he used to guide me through things beyond physics. His main advice was always to maintain absolute integrity in both academic and non-academic matters. It was his expectation from every other colleague, including his students. \\

The last time I met him was in February 2020, when he visited IIT Gandhinagar for an institute colloquium. He was so full of energy with many intriguing research ideas. Also, he was planning for an undergraduate-level physics lecture course. His premature departure was such a sudden shock to all of us. It is still difficult to reconcile his absence among us. Nevertheless, his intellectual achievements will always be there to guide and inspire generations to come.

\bigskip

\bigskip

\centerline{{\bf Valerio Faraoni} ({\it Bishop's University, Sherbrooke, Canada})}
\medskip
\noindent 

The news of the untimely passing away of Professor Thanu 
Padmanabhan  on 
September 17, 2021 was a shock to all 
of those who knew him. Paddy was an icon of the Inter-University Centre 
for Astronomy and Astrophysics in Pune, India. Researchers working in 
theoretical gravity, even those who did not know him personally, have 
encountered multiple times Paddy's work, his textbooks have occupied the 
study nights of countless students, and his popular science articles have 
inspired and delighted many readers. Even without knowing Paddy 
personally, one 
cannot help being impressed by the depth and the sheer amount of his work. 
This was the case for me when I headed to IUCAA for one year as a young 
postdoc in the late 1990s. There were many factors attracting me there, 
and certainly one was the fact that the famous Professor Padmanabhan 
worked there. Because of his reputation, his renowned works on gravity and 
thermodynamics, and his remarkable research output, I was expecting a very 
busy man somehow difficult to approach. Nothing could have been more 
wrong: busy, he was, but he was also very approachable. It did not matter 
who you were or where you were coming from, student, postdoc, or senior 
professor, Paddy would listen to you if you had something pertinent to 
say, and he would encourage opinions and interaction. I felt  
that he was always looking for a different point of view, a different 
angle to view a problem.

My first impression was that discussing with Paddy meant that you really 
had to focus on the subject. More than once I said something that sounded 
stupid to me as soon as I said it and realized that Paddy had already 
worked out something similar in a paper with much more clarity and vision. 
But he would always listen anyway and never point out the triviality. He 
was very deep and such discussions could not be superficial, one was 
forced to think hard and realize the limitations of one own's imagination. 
Paddy loved the topics he worked on, especially the thermodynamics of 
gravity and emergent gravity, and he would get excited if a new glimpse 
would promise to peep out. He was also very passionate about astrophysics 
and cosmology, semiclassical gravity, and spacetime singularities.

Technically, I was not Paddy's postdoc at IUCAA but I appreciated very 
much the interaction I had with him. Many years later, he kept sending me 
his new papers on the thermodynamics of gravity that he was excited about 
and I regret I could not find the time to understand them completely and 
interact more with him. As much as I kick myself now, I 
was not working on these topics at that time and to make progress on 
Paddy's new trail, or just to absorb it, required a committment of 
time and focus that I could not put in at that time. I must have 
disappointed him on that. I am also sure that he understood.

Re-reading Paddy's papers, I can't help thinking that there was more 
underneath, as if he was somehow visualizing in his mind the emergence of 
gravity, although eventually a visual picture cannot be contained in those 
mathematical formulae.  I believe that these mental images and this 
visualization process still live in the minds and the work of Paddy's many 
students and collaborators. Thank you Paddy, I am glad that our light 
cones intersected.

\bigskip

\bigskip

\centerline{{\bf Sunu Engineer} }
\medskip
\noindent Paddy began as a teacher to me and in many ways defined the concept. Years before we met in person, through the columns that he wrote in the science magazines. I grew up with them, learning, being challenged, enjoying, but not knowing the source. It was years later, after I came to IUCAA that this charming fact made itself apparent. 
\par 
Being his student was a good but hard journey. While it  was not in his nature to go easy on a student with respect to work and its quality, he was a good friend, a great teacher and story teller with an incredibly wide set of interests and grasp of the world beyond much of his published works. We shared a certain cultural perspective along with a language and many are the insights that he directly or indirectly seeded in our continuous interactions even a day before his death. Most of his  early students and their families  were also close to Vasanthi and Hamsa, Paddy's family. It was in all ways a traditional 'Gurukulam'. Which accounts for why I speak for all of us, when I say that the news of Paddy's passing away created an instant vacuum in our lives on many fronts. To say that he is missed would be a cosmic understatement.
\par
The fluid versatility of his mind that shifted through ideas,  astronomy, artificial intelligence, physics, quantum computing, chess, consciousness, COVID, stories, history, both broadly and deeply at the same time, and at high speed, was often taxing and many conversations with him can be characterised as 'delayed enlightenment'. He would be happy to know that he was in many ways Krishna to his students, guiding them in battle and otherwise.

\bigskip

\bigskip

\centerline{{\bf Sumanta Chakraborty} ({\it Indian Association for Cultivation of Science, Kolkata})}
\medskip
\noindent Let me ``backtrack for a moment" (This was one of the most used phrase by Paddy) and recall that Paddy had called me on 16th September in the evening as he was worried about some calculations I had shared with him. This has to do with the ``River Metric", which has the ADM lapse function as unity and the spatial metric to be $\delta_{\alpha \beta}$ (Paddy was always fond of the convention in which spacetime indices are denoted by Roman indices and spatial indices are denoted by Greek indices. This is possibly connected with a story that Kandu had shared, in which Paddy ran out of Greek indices and started using Malayalam letters!), with only the shift vector $N^{\alpha}$ being non-zero. The extrinsic curvature and its trace are dependent only on the symmetric combination, $(D_{\alpha}N_{\beta}+D_{\beta}N_{\alpha})$ and since the spatial metric is flat, the Ricci scalar is also dependent only on this symmetric combination. However, seemingly in the literature, the Ricci scalar has been expressed as a quadratic quantity in the antisymmetric combination $(D_{\alpha}N_{\beta}-D_{\beta}N_{\alpha})$. Paddy was rather disturbed, as neither him, nor me could resolve this issue over our discussion through phone for the last couple of days of that fateful week. On 16th September, I thought I found an explanation, but Paddy was not convinced (In my email including the calculation, I had suggested that possibly this will make the confusion go away and Paddy had responded with an email, whose subject line says ``no", which is the last email conversation with him.). However, after discussing this issue with Paddy over phone, I sort of understood where the real problem is lying. Then we both agreed to look at the spatial part of the Einstein's equations to resolve this confusion. 

On 17th September, I was doing this calculation and was expecting a phone call from Paddy (Getting phone calls from Paddy was a part of my life, whenever he was unhappy about some calculations, have some new Physics ideas, he faces technological problems with his iPad and/or software installation, he would call me right away.) regarding the issue mentioned above. Instead, I suddenly got a call from Sriram and I do not recall exactly what he said, except that I understood that I will never ever get any further phone calls from Paddy in my life. I could not believe myself!

This unbelievable and untimely passing away of Paddy has been a tremendous loss for me. As my grandfather used to say, ``finding a proper friend, philosopher and guide is the most difficult thing in one's life'' and I believe myself fortunate enough to have found my guiding star, Paddy. But at the same time, while writing this article, I feel very unfortunate that I will miss innumerable occasions to learn from Paddy, following his untimely demise. I think interactions with Paddy have been some of my most enjoyable and memorable as well as intense learning moments. I strongly believe it is not only difficult but impossible to find a better supervisor than him. Despite being extremely busy, he heard my problems with patience and with his amazing insight gave advice, which in most of the cases resolved them. Most importantly, he gave me complete freedom in doing research, except for teasing me in his unparalleled ways for working on alternative theories of gravity and gravitational waves. 

Another aspect of Paddy needs to be highlighted, namely, his respect for fellow researchers. As I joined Paddy, I told him not to call me after 9 PM as I go to sleep early and he adhered to this throughout our collaboration. Whenever he calls me up after 8 PM, he will always ask, ``Are you asleep? Can we talk now?", a very kind gesture from such an internationally acclaimed scientist. Such a personal touch in our discussions will always keep him alive in my memories for years to come. Not just Paddy, Vasanthi and Hamsa are also very kind and we all had various socio-economic discussions over lunch and dinner, where I also had the pleasure of enjoying the tasteful dishes prepared by Vasanthi. I will cherish all those glittering moments throughout my life. 

I do believe that wherever Paddy (Paddy: Prof. T. Padmanabhan, ADM: Arnowitt-Deser-Misner, Sriram: Prof. L. Sriramkumar, Kandu: Prof. Kandaswamy Subramanian, Vasanthi: Dr. Vasanthi Padmamanabhan and Hamsa: Dr. Hamsa Padmanabhan) is now currently, he must be working on new Physics ideas with excitement. He may also be playing chess with god and teasing Einstein, ``You are correct, god does not play dice, but god does play chess", in his inimitable manner.   

\bigskip

\bigskip

\centerline{{\bf Krishnamohan Parattu}  ({\it Valparaiso University, Chile}) }

\medskip

\noindent Paddy's demise was sudden and unexpected. Being the same age as my father, Paddy was a father figure for me. I fondly remember the many dinners my wife Sonia and I have had with him, his wife Vasanthi, daughter Hamsa and his other students and postdocs. It felt like I had another family at IUCAA. After coming to know of his passing, the next couple of hours were spent reading whatever I could find about him… recent interviews and posts by people in social media. Then, not knowing what else to do, I wrote whatever was coming into my head as a poem in his honour, shedding more than a few tears along the way.

A week before, he had called me in connection with an old project and asked me for a reference that I had earlier mentioned to him.  After suggesting some calculation with respect to that project, he briefly spoke about some of the other projects he was working on. I later found and sent him the reference but could not find time to do the calculation that week. For that, I felt guilty.

While still in school, I was aware of Paddy due to interviews that used to appear in newspapers in Kerala. I came across his website when I was an undergraduate. His wide interest in theoretical physics resonated with me. Later when I was thinking about my PhD, after having set the criteria as research area, guide and institute, in that order, and having set quantum field theory or gravity as the values the first variable can take, Paddy was the name that naturally popped into my mind. Thankfully, things worked out and I was able to join him.

Paddy had a no-nonsense but liberal approach to mentorship. He said once, ``I will not tell you what to do. I will just tell you that if you do this this is going to happen."  Even though his main research focus during the last decade was on the emergent paradigm, Paddy used to give his students a variety of problems and then work on the ones that they got interested in. When I joined, he offered me a choice between emergent paradigm and a project in structure formation that involved a code. In the case of Sumanta, I heard he had offered the options of cosmology, astrophysics, general relativity, etc. and emergent paradigm along with the likelihood of obtaining a good postdoc post PhD in each. Emergent paradigm had the least likelihood. Both Sumanta and I chose emergent paradigm. 

In the fourth year of my PhD, I informed him of my intention of going to a conference. He said it is better to go in the final fifth year. But he further said that this was the same advice he gave to my senior Sanved and Sanved had still gone for a conference earlier. I decided to go, and he helped me with writing to the organizers for financial support. In that case, it turned out to be the right decision. The conference gave me a broader view of what is happening in the community, and I also discovered research in a direction very close to what Paddy and I had been doing.

He believed that, while PhD was his responsibility, students should make their own way after that. He used to say, ``At your age... if you had gone for Indian Administrative Service, you would have been in charge of a district! Once, admitting that his recent students have had issues finding a good postdoc right after PhD, he said with conviction, ``You are all good people. Once you are out there, you will do well!''

Though serious about his work, Paddy used to keep things light and fun. There was this period just before our marriage that Sonia used to come and visit me in IUCAA. The people in the administration were tying themselves in knots worrying about a non-academic person coming to visit regularly. Finally, an admin person went to Paddy and informed him. Paddy quipped, ``Just make sure that it is the same girl every time!'' Paddy had already met Sonia and so he proceeded to reassure the admin that it was alright.

When I think of Paddy, what I remember most is the infectious enthusiasm he used to exude. It was difficult to imagine that he ever slept, and even more difficult now to come to terms with the fact that he has gone for the eternal sleep. I end by paying tribute to a great teacher, mentor and friend.

 \bigskip
 
 \bigskip

\centerline{{\bf Mohammad Sami} ({\it SGT University, Gurugram})}
\medskip
\noindent {\it Memories and impressions of T. Padmanabhan}

\noindent Professor T. Padmanabhan was one of the pioneers of Indian Theoretical Physics, who  played an important role in training  young researchers through his lectures, research reviews, and excellent books. In the Indian context, he was the ultimate person to consult if there was confusion about the general theory of relativity, cosmology, or field theory in curved space-time. Obviously, Paddy’s sudden disappearance is an unimaginable loss, especially to the Indian physics community. He has left a rich research and academic legacy in terms of his research contributions, lectures, reviews, books he published and students he trained. His academic support to researchers working in Indian universities was phenomenal. He played an important role in building the Centre for Theoretical Physics at Jamia Milia University, New Delhi. My first and direct encounter with Paddy happened in 2003 when I was a long term visitor at IUCAA. He suggested to me a particular form of coupling of dark matter with dark energy and asked me to check the viability of the model. A few days later he left for Cambridge for two months. I worked on the project and obtained the expected results but did not write to him. Paddy was extremely systematic and disciplined and I am otherwise. One fine morning when I met him after his return, he sounded a bit unhappy as I had not corresponded with him about the project. After looking into his mail box, he called me and said: “you are impossible, I am reading the draft and will get back to you in the evening”. I remember, we exchanged mails from six p.m. to two a.m. when he finally approved the draft and I then submitted the paper to  the arXiv. Actually, he could have been the coauthor in several papers of mine but just asked me to acknowledge him.

    My stay at IUCAA from 2002-2005, were the most active years of my research career as a person like Paddy was around from whom I learnt a lot. Whenever I was stuck, I would call Paddy and he was always kind enough to explain to me on the black board keeping the level of discussion such that I could understand; he knew that I did not have a cosmology background. He support to me for building CTP at Jamia was phenomenal; he was  the chairman of the review committee of CTP. 
    
         Four days before he left us, we had a long telephonic conversation which was unusual of Paddy as he normally used to be brief. He promised me to lend his full support to the new centre for Cosmology and Science Popularization (CCSP) whichI am trying to build at SGT university, Gurugram. 
         
Paddy will live in our memories, he will live through the  academic and research legacy he has left behind.

\bigskip
\bigskip
\centerline{{\bf Sandipan Sengupta} ({\it Indian Institute of Technology Kharagpur})}
\medskip
\noindent {\it Remembering Paddy}

\noindent As a postdoc at IUCAA (during 2014-15), my interactions with Paddy were little, but were
sufficient to glean the fact that Paddy was his own man with his own ideas. He was one of the very
few who could offer new perspectives from a technical as well as intuitive viewpoint while
discussing a problem or a paper. That he was creative was no secret, but those little one on ones
over the blackboard in his room while trying to convince him of an idea perhaps emerging afresh in
the literature, or while trying (mostly failing) to argue that a particular point made in his own paper
looks suspect, would almost always end up in me coming out with an urge to either look up
something new which I had not thought about, or to dig into an interesting paper which I had not
read, or to learn a technique I did not know, or at least come out with a clarity as to whether one was
treading the right path of thinking in his view. This was remarkable.

I have always enjoyed learning from him from whatever little opportunity I have had during my
shortish stint at IUCAA. Our very last discussion was barely a few months old, concerning the
problem of cosmological constant. And as usual, the series of (email) exchanges over a couple of
days ended in a little but a fresh bit of clarity, and that is the least Paddy would always do in an
academic discussion. I would value that, and more than anything else, I would miss that from now
on, too, along with the warmth of it.

\bigskip
\bigskip

\centerline{{\bf Bibhas Ranjan Majhi} ({\it Indian Institute of Technology Guwahati})}
\medskip
\noindent {\it Post-doctoral fellow}

\noindent 
Being a M.Sc. student of physics I first heard the name, Prof. Thanu Padmanabhan, a great mind in theoretical physics. At that time I was not familiar with his works, but at least it was known to me that he is trying to illuminate ``quantum mysteries of black hole''. Although not fully understood, but his few articles attracted me then and later inspired me to take up physics as a profession. I joined S.N.Bose national centre for Basic sciences, Kolkata as a Ph.D student and started working on gravity. Hawking effect \cite{Hawking:1974rv,Hawking:1975vcx} and black hole thermodynamics \cite{Bekenstein:1973ur,Bekenstein:1974ax} were my interests and incidentally I came across several related articles of Prof. Padmanabhan. His way of explaining the things based on the basic physics is always reflected in his articles. Compact and lucid presentation give me joy when I go through them.  Among these the papers on the understanding of Hawking radiation via tunnelling approach through Hamilton-Jacobi formalism \cite{Srinivasan:1998ty} attracted me a lot and helped to define my Ph.D thesis problem.  Incidentally, my thesis \cite{Majhi:2010onr} was on different aspects on tunnelling approach to understand black hole  thermodynamics. During that time I had several questions in my mind related to this topic and was always eager to discuss with him.

I physically met first him in a conference (year 2009) at Saha Institute of Nuclear Physics, Kolkata. Being a new comer in this area I was afraid to face him, but was always excited to talk to him. Somehow I managed to catch him and started telling about my views on the tunnelling approach. It was a life time experience for me. His way of explaining the topic and attitude was something to watch. In between the conversation, I mentioned that in tunnelling approach one calculated the tunnelling probability of escaping a particle from the black hole horizon which incidentally comes in the form of Boltzmann factor. This only determines the temperature of the horizon. Whereas Hawking was able to show the radiation spectrum. I asked him whether determination of such emission spectrum is possible within tunnelling formalism. He quickly commented that ``it is not clear to me how to proceed, rather may not be possible as Hawking's calculation was based on quantum field theory.''  Later I was able to calculate the emission spectrum from horizon in tunnelling formalism \cite{Banerjee:2009wb} and dared to communicate with him. He was pleased as his approach can overcome the foreseen apparent limitation. In reply he mentioned to me several rules to grow up as a successful physicist. That meant a lot for me at that time and I realised that he is not only a great scientist but also a great mentor for us. 

His brilliant idea of assigning temperature and entropy on a local horizon (not necessarily black hole) gives rise to a clear connection between gravity and thermodynamics. He proposed an idea that the gravitational dynamics may not be a fundamental theory, rather it  emerged from an underlying microscopic theory of thermodynamics \cite{Padmanabhan:2014jta}. This novel idea, based on semi-classical approach, shows a different way of understanding the mysteries of gravity. After completing my Ph.D, I went to IUCAA in February, 2011 as a Post doctoral fellow. I was very excited to work under Prof. Padmanabhan. First day I went to his office to meet him. I was little bit scared and always thinking about how to speak to him. We had just few words on that day. He asked to meet me later after finishing all the official formalities of my joining at IUCAA. After two days when I was in my office room, he came to me and asked me whether I will be able to discuss then. His first question was --  ``what is your plan of work?'' I was little nervous and was not sure whether my proposal will sound interesting to him. I was silent for some moment. Then looking at me he said that ``you may start from zero''.  That single sentence blew my all fear and drew probably the most defining moment of my life. Probably he had a great sense of making things easy for a student and believe me that  boosted me enough confidence on my ability. I informed him that ``I am interested in emergent paradigm of gravity; particularly whether the nature of degrees of freedom, responsible for horizon entropy, can be illuminated through Brown-Henneaux-Carlip's \cite{Brown:1986nw,Carlip:1998wz} Noether charge and asymptotic symmetry approach''.  

He was very much delighted to hear this and immediately sent me few references to look at and kept sending relevant literature almost every day. His devotion and continuous encouragement helped me to understand the whole background of the research and probably played the most important role for what I achieved today. The whole study  consisted of few parts -- (i) definition of a suitable charge corresponding to the diffeomorphism symmetry of the spacetime, (ii) identifying a proper definition of bracket among the Fourier modes of this charge and finally (iii) identification of suitable diffeomorphism parameters. His insightful comments and suggestions indeed helped to overcome all these steps very successfully. In this regard it may be pointed out that his earlier study \cite{Padmanabhan:2009vy} on gravitational action revels several thermodynamics aspects of horizon. He showed that the surface part of the gravitational action carries all the information of the bulk part and interestingly calculation of surface part on the horizon yields entropy of the horizon \cite{Padmanabhan:2009vy}. Motivated by these findings he was very much curious whether Noether charge corresponding to surface part of the action has anything to do with horizon entropy. We found that it is indeed 
true and well applicable to any generic local null surface \cite{Majhi:2012tf,Bhattacharya:2018epn}. 

Throughout my time with him, I felt that he has immense confidence on his students and that helped to grow up the self confidence in us.  My last meeting with him was on August, 2021 over phone. We were doing a project on the thermodynamics of a generic null surface. He was so enthusiastic  about the scientific discussion that he always called me and informed about several possible directions of thinking. I was so habituated with his often phone calls that without this my day does not complete. Why not? After all it is a great opportunity for me to learn several things from him. Now I will miss it Sir. 

During the conversation, I always call him as ``Sir'' and I found that he was not so comfortable with this exhortation. One day he told me to call as ``Paddy''. It was really difficult for me to call a person like him by his nick name and therefore I told him that ``I will try.'' I can still remember a pinch of smile on his face after hearing this. Truth is for the rest of his life, I never called him as ``Paddy''. I am sorry ``Sir'', it is not possible for me and please be ``Sir'' for probably only one member of your Paddy family -- miss you Sir.

\bigskip
\bigskip

\centerline{{\bf Kinjalk Lochan} ({\it Indian Institute of Science Education and Research Mohali})}
\medskip
\noindent {\it Thanu Padmanabhan: A life - Horizonless}

\textit{How does it feel to be working with such a big-shot as Paddy ? He is such a big star ! } This  was asked to me by a  fellow participant at a \textit{Marcel Grossmann meeting} in Rome, July 2015,  where I had presented a work that was  just completed in collaboration with Paddy. I did not know how to answer that, but it got me into thinking. I was of course in full knowledge of the respect, repute and aura Paddy commanded. I was also in complete awe of his mesmerizing brilliance, insightful knowledge-base and above all the enthusiasm in breaking things down to the basic fundamentals, where he could see each and every brick of the construction. He was not to be content just with answer to a query. With child-like curiosity, he would break things down and reassemble them to see how smoothly it works. What got me thinking was the usage of the  phrase of \textit{big-shot}. About two years back in 2013, when I was moving from TIFR to join the research group of Paddy, I also had a recognizably similar feeling and perhaps a kind of trepidation in going to work with a \textit{big-shot}.  Still here I was, trying to remember -- how and when this diffidence  had faded away in the warmth, self assurance and affinity, this father figure offered. I had read from youngsters joining Indian cricket team,  Sachin Tendulkar the unchallenged master of the game, would not ever let them feel intimidated by his stature. I guess geniuses have a common trait. Still, one simply can not box Paddy to just a genius, he was much more than that.

Apart from being known as the renowned researcher and prolific speaker that he was,  Paddy was an excellent human being. Consistency, sincerity, eagerness to learn from everything and everyone, joyfulness, humour  and to top these-- humility, there was no discussion where colors of all these attributes would not be radiantly visible. I can not imagine a more apt personification of  \textit{Rishis} in modern times,  perfecting the practise of \textit{Yoga},  \textit{Dhyaan} and \textit{ Manan}, revealing the outcomes of their musings  in the treatise to follow the churning. Many would agree that his was more of a \textit{Gurukul} than a mere \textit{Group}. Anyone entering this \textit{Gurukul} would assuredly feel a part of the family environment,  which Paddy, Vasanthi and Hamsa had so carefully and thoughtfully crafted, melting all inhibition one would enter with.
 
Paddy's philosophy of doing science was simple yet elegant -- \textit{ ``for the fun of it''}. I am aware almost everyone in the arena of research would proclaim this as their driving cause, but couple this with the untiring, uncompromising earnestness that Paddy had, and you will feel the conviction behind these words. For last few years I had the privilege of discussing physics (along with society, literature, languages, films,  life and much more), I can not find one empty statement which was to be later found in contradiction to his actions or this philosophy. Often I wondered what keeps him driving with such an unending energy.  The answer, I understand, is in being absolutely truthful to what one believes in, and that most certainly he was. Even close to the \textit{bureaucratic retirement}, he was at his absolute prime, playing like a maestro till his very last. These past few months, we kept talking to each other frequently for longish discussions related to a collaborative work, he was apparently so much in fun with. I was supposed to call him in the afternoon of the fateful day, to discuss the revision of what now remains as his very last research work, available in  public domain.

So I agree with that fellow participant of the meeting. Paddy indeed was a gigantic, magnanimous star. Very bright, brilliantly shining and  an absolutely unique one. He shined with his own certainty, acting like a guiding star, providing definitiveness to the quest of direction, simply by its presence. It would not forcefully drag you to itself by virtue of its immense gravity, but would rather shine brilliantly, gently and with a profound calmness.  In older times, sailors in the deep and the dark of the oceans would use  some guiding stars in the sky to orient themselves, if they so require. I believe such is the case with research, many stars in the sky but very few, to guide you through.

On the morning of his demise, I was invigilating an exam and was about to discard the call which broke the heartbreaking  news that our guiding star had bade adieu to us all, as quickly and silently as it had once appeared on the horizon. Night sky had hushed into a lull, consuming what shone so brilliantly for so many decades. As for us, who had the chance to orient our sailings in his breeze, we grieve together, yet it is imperative on us to not let go of the sight of the quest or directions his radiance illuminated.

\begin{figure*}[!h]
\centering
\includegraphics[width=12.00cm]{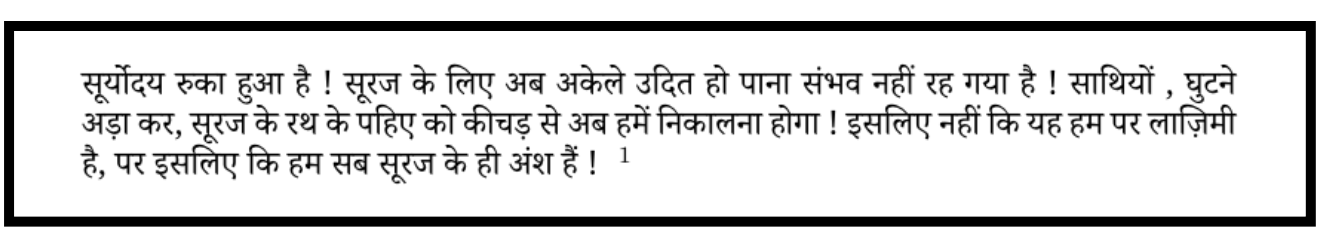}
\caption*{Dharmvir Bharti 
quoting from \textit{March of the Spirit} by Angelos Sikelianos}
\end{figure*}

\noindent \textit{"The dawn has got stuck in its course. It is not possible for the Sun to rise alone anymore ! 
 Forward friends, push with knee and chest to get it out of the mud !
 Not just because it is on to us to break the day, yet because we are all part of the Sun"}
 
\bigskip

\bigskip

\bigskip

\centerline{{\bf Suprit Singh} ({\it Indian Institute of Technology Delhi})}
\medskip
\noindent {\it \underline{paddy — all in lowercase} }

\noindent they say, once something is inked it is pakka (even paddy used to remark “…the argument is pakka”). may be that’s the reason i have been very reluctant in writing this note. it’s hard to have it seep in that paddy is no longer with us on this planet we have all come to call home. the question, therefore arises, is it? to many this question would be useless to ponder upon. not paddy though, he had exclaimed with the quote first line on his now forever frozen webpage, “Beam me up Scotty! There is no intelligent life out here.”  paddy was a sculptor of one of the best kinds. the way he has shaped many of his students lives is a proof of his excellence at that by itself.  there were many aspects of the grandmaster. i will list some of the small tidbits here from my personal collection. 

paddy was a great player of chess so much so that when he was young, at his village with a friend of his,  he used play chess, without any pieces, mentally and at the end, they would shake their hands and leave.

paddy did not like sweets. well, he loved chocolates, and particularly was fond of my mom’s homemade chocolates.

paddy was also extremely fond of watching movies and yeah he watched game of thrones too like many of the millennials, on a projector in his weekend home. he told me he used to work while savouring those episodes.

paddy drove an automatic car volkswagon vento; but when he had a maruti with a manual gear box, he said he would shift it to the second gear and drive it also like an automatic. 

paddy and his family knew my family very well. my sister who is a chartered accountant, was pursuing the course at that time, and paddy once remarked to her, “… getting a phd is  a much easier job than becoming a chartered account”. 

paddy’s emails were written like this — in lowercase — all the characters, words and phrases and many lacked punctuations, to him, all that mattered was the content

paddy used command-line pine for his email management. only recently when pine become too complicated due to protocols etc he asked me about alternatives and i told him about mozilla thunderbird.

paddy’s whole life was a class act. i once asked him about how does he decide what to wear on a day, he told me he would just pick up whatever was on top of the pile

paddy’s forever now webpage also contains “the answer” in his lesser known write-ups on the philosophy of life, upanishads, and the bhagvada Gita. 

i am reminded now of a wonderful quote from j k rowling, through the character of albus dumbledore, “However, you will find that I will only truly have left this school when none here are loyal to me. You will also find that help will be given at Hogwarts to those who ask for it.”  to paddy this loyalty would be best displayed through perseverance and excellence, and doing things {\it just for fun}.

\bigskip

\bigskip

\centerline{{\bf Aseem Paranjape} ({\it Inter-University Centre for Astronomy and Astrophysics, Pune})}
\medskip
\noindent Words alone can never do justice to the phenomenon that was Paddy. But he has left us now, so words and memories will have to suffice. I can only hope that this (highly incomplete, very personal) account of my association with Paddy will play some small part in keeping his memory alive in the days to come.

I first encountered Paddy when, as a wide-eyed undergrad, I was selected for a summer project under his guidance. Not surprisingly, Paddy's reputation for being a tough taskmaster preceded him and left me quite apprehensive upon seeing the words 'You will work with Prof. T. Padmanabhan' in my invitation letter. Upon reaching IUCAA, I was convinced by Harvinder and Jasjeet, who were visiting at the time, that there was no need to worry because 'Paddy is a really sweet guy'! Indeed, I was soon treated to the full Paddy experience in his air-conditioned office, quickly moving from niceties to a rapid physics interview judging my preparation, to a proper lecture on relativistic actions, complete with the famous left-to-right chalkboard switcheroo. Not yet sure about the 'sweet' part, I did leave the room with a strong feeling of anticipation and excitement for the coming two months. A week into the project, I was hopelessly stuck at some point in Landau-Lifshitz classical theory of fields. In desperation, I recalled Harvinder and Jasjeet's advice and, heart-in-mouth, called Paddy at home on the telephone from my guest house room. I explained how stupid I was feeling because I was unable to derive the Einstein equations from the Einstein-Hilbert action. After asking how I had chosen to proceed, he quickly dispensed with my unnecessarily complicated approach (involving the Euler-Lagrange equations with tensors) and told me to open out a few terms and then 'just vary the action, okay?'. The expert reader will appreciate the elegance of this approach, while the non-expert can try to imagine the relief and awe a young student such as I must have felt when Paddy so kindly held out his finger to deftly reveal a beautiful and simple path towards a more complete understanding of the elegance of general relativity.

This sharpness of intellect and clarity, combined with a unique and almost oxymoronic blend of generosity and no-nonsense straight talk (the kind cruelty of the surgeon's knife), was a hallmark of Paddy that I repeatedly experienced over the course of almost two decades. When, after a field trip to GMRT, I was confused about whether to pursue theoretical physics or radio astronomy for my Ph.D. at TIFR, it was Paddy's talking-to (this time over telephone from IUCAA to my home in Borivali) which convinced me to continue developing theoretical expertise. Throughout my subsequent Ph.D. with T. P. Singh, Paddy was my 'other advisor', hosting me at IUCAA periodically for several short visits which invariably super-charged my enthusiasm for not only my work with him and for my thesis at TIFR, but for physics in general. As a postdoc in Italy and then in Switzerland, I experienced a different side of Paddy, walking the streets of Rome and Zurich with him and Vasanthi, enjoying coffee, pizza and hilarious anecdotes, on their short but thoroughly enjoyable visits. Interesting events seemed to simply follow him around. As a brief example: Paddy's wallet was picked by a gang of thieves in an elevator in the train station in Rome, and thereafter recovered from the railway police (sans cash), in a manner which he conceded was clever enough that the thieves deserved the money they got! I was also lucky enough to become part of a 'Paddy drinking milk' story at a cozy little restaurant in Trieste with some dear Italian colleagues.

At IUCAA as faculty, I saw yet another side of Paddy. No longer a student or a postdoc myself, I had wondered how my relationship with Paddy might evolve. My scientific interests had meandered over the years and had drifted into cosmological large-scale structure, somewhat far from my early fascination with classical general relativity (whose connection with thermodynamics remained Paddy's focus). Once again, I needn't have worried... Paddy smoothly and graciously transitioned from academic mentor to senior colleague, offering support, encouragement, criticism and advice on myriad administrative and academic matters that were every bit as sharp and effective for a young faculty as had been his guidance on varying actions for a young student way back when. Paddy's thought-provoking and totally original contributions to INAT questions (IUCAA's Ph.D. entrance exam) will be sorely missed! And I can honestly say that my most memorable and intense experience at IUCAA so far has been the organisation, along with Tirth, of a celebration of Paddy's 60th birthday by bringing together many of his closest academic friends for a conference. The feathers we ruffled in the process were, in my opinion, well worth the outcome! 

During these last several years, Paddy being Paddy, of course did not waver from periodically ribbing me about not working on 'tough problems'. In turn, I also tried to rekindle his interest in large-scale structure, not least by asking him to serve on my student Sujatha's 3-member advisory committee (to which he agreed without hesitation). I cannot say that I was fully successful in my endeavour, although I did manage to get him to spend several hours discussing relativistic effects at ultra-large scales. 

It hurts that I will no longer have another opportunity to try (even unsuccessfully) to alter his unwavering, ramrod scientific focus. It hurts that I will not speak to him any more, nor wave hello while on a run as he passes by on his evening walk with Vasanthi. It hurts that his voice and laugh won't echo any more across the Foucault pendulum and into our offices. His memory will endure, though... through work, through physics problems, through his books. Through these, perhaps, his voice and laugh might yet echo once more in our minds.

\bigskip

\bigskip
\bigskip
\centerline{{\bf Krishnakanta Bhattacharya} ({\it Inter-University Centre for Astronomy and Astrophysics, Pune})}
\smallskip
\noindent{\it Academic and personal interaction with Paddy}

\noindent  I used to know “Paddy” since my graduation by watching his YouTube lectures and used to admire him like a distant star whom I could behold but remained too far from my reach. After entering into PhD, I heard many interesting stories about him from my PhD supervisor Dr. Bibhas Ranjan Majhi, who was also a postdoc fellow under Paddy’s mentorship. At that point of time, I used to think how lucky those researchers are who have got the opportunity (rather earned) to work with him and have seen him from the close corner. Needless to say, when I was about to complete my PhD and was searching for postdoc positions in various places, I had also applied to IUCAA hoping I might also get the opportunity to work with him as a postdoc fellow. Thus, when I was selected, I was ecstatic; finally the distant star seemed well within my reach. I had denied several offers in India and abroad in order to work with the genius of my research area. I can still vividly remember my first conversation with Paddy which incidentally happened in a telephone call because of the pandemic, finally I could hear my hero suggesting me with his extraordinary insights. Initially I was in an awe of him but he made me feel very comfortable. He had asked me to record all our telephone conversations for my convenience. Unfortunately, due to the pandemic, most of our discussions took place through phone calls. During each and every conversation, I was enriched with his brilliant insights and wonderful ideas. I rather feel myself unfortunate who got the opportunity to work with him for a very little time and for being (probably) the only member of “Paddy’s Gharana” who have no publication with him. One of our work, investigating the consequences when the constant parameters of the metric are promoted to the spacetime dependent variables, was in progress. We got some initial results but, he wanted to dig it further to make it worthy of being published. Apart from this direct work with him, the discussions with him motivated me to investigate several other aspects of gravitation.

    Apart from being a wonderful mentor, who was rather strict in terms of completing work and about deadlines, I have noticed a humane side of him when he got to know about the fact that my father is a chronic-kidney-disease patient and undergoes dialysis twice a week. He had assured me that he can make all the arrangements so that I can work from home whenever required. It would help me to take care of my father and to continue research work simultaneously. Just two weeks before his demise, my father again fell sick due to a stroke and urinary tract infection. As soon as I had informed him about it, he told me to go home and concentrate upon the medical emergency of my father, whereas he would take care of all the formalities once I return back; hardly had I known that when I shall be able to make my journey back to IUCAA, he would have departed for the heavenly abode.

     I remember, after getting selected in IUCAA as a postdoc fellow, I had sent him a mail asking him whether he would stay at IUCAA during my whole postdoc tenure, as I knew that he was going to retire soon.  In response, he had informed me that he would be in IUCAA for the majority of my postdoc tenure and, even if he retires, he would stay in Pune. Thus, I would be able to collaborate with him for my entire postdoc at IUCAA. While he was a man of his word but, unfortunately, he could not fulfil his promise regarding collaboration with me and I consider myself very unfortunate to be mentored by him for a brief period of time. However, I shall cherish all those moments which I have shared/discussed with him and shall treasure all the recorded telephone conversations for my whole life.  I shall miss my postdoc mentor very badly. May his soul rest in peace.

\section{Paddy as a role model teacher: a short story}
\noindent {\it On Feynman's formula for the electromagnetic field of an arbitrarily moving charge,
American Journal of Physics, 56, 1036 (1988)} [A.R. Janah, T. Padmanabhan and T. P. Singh]

\noindent This is a heart-warming physics story from the time when Paddy was at TIFR [1979 - 1991].
We all knew what an extra-ordinary teacher he was. The tidiness of the black-board use, the elegant hand-writing on the board, and the clarity of the presentation. Almost never he used notes while teaching - the day's lecture, from start to end, would be worked out on the board, as a piece of inevitable logic. He constructed equations on the spot, from first principles.

In 1987, Paddy taught a one-semester course on Classical Electrodynamics. As his then Ph. D. student, T. P. Singh also helped as a Teaching Assistant for this course, his job being to prepare problem sets (in consultation with him) and to check the answer sheets.
Every assignment would include one open-ended, non-textbook style, problem. e.g. under what conditions would you see a full circle of the rainbow, after it has rained and the sun has come out?

In his Lecture Volume II on Electrodynamics, Feynman gives an elegant formula for the fields of a moving charge [Eqn. (1) in above paper] but he does not give the proof, leaving it for the reader to figure it out. Neither Paddy nor the tutor knew how to prove this formula (it is non-trivial, the proof), but they gave it as an assignment problem.
After handing over the assignment, they struggled for many days to prove it. They could not. The students turned in their answer sheets after two weeks, nobody could prove it, but one student A. R. Janah made good progress.

The students wanted to know the proof, which the teacher and tutor did not have, and they were still struggling. One evening while Paddy and the tutor were
at his home Paddy had a brilliant insight [you will see it in the above paper] which then quickly led to the proof.
This find was important enough to be published because to their knowledge the proof was not to be found in the known literature. 

Now for the heart-warming part of the story. This published proof is entirely Paddy's doing. But he insisted that Janah and the tutor also be co-authors, because they had tried hard.
The paper was published in American Journal of Physics in 1987, and they were happy to learn that later Am. J. Phys. listed it in its Top 100 (or something like that) memorable papers. You will see the hallmark Paddy style clarity of logic in it. Several physics teachers wrote to Paddy appreciating that this paper filled a mischievous gap which Feynman left in that chapter.

Many evenings would be spent at Paddy's home by his students, calculating, staying back for dinner with the family, including his very affectionate parents. His father would ask them jokingly if Paddy was treating us well or should he give Paddy a scolding?  Food was very typical south Indian fare: rice, sambhar, south Indian veg, and Paddy's favourite pappadam. Every meal had to have pappadam.
It was a cheerful gurukul atmosphere [but very strict in the office].

\section{The 50th and 60th birthday books, and the 60th birthday conference} 
\begin{figure*}[!h]
\centering
\includegraphics[width=12.00cm]{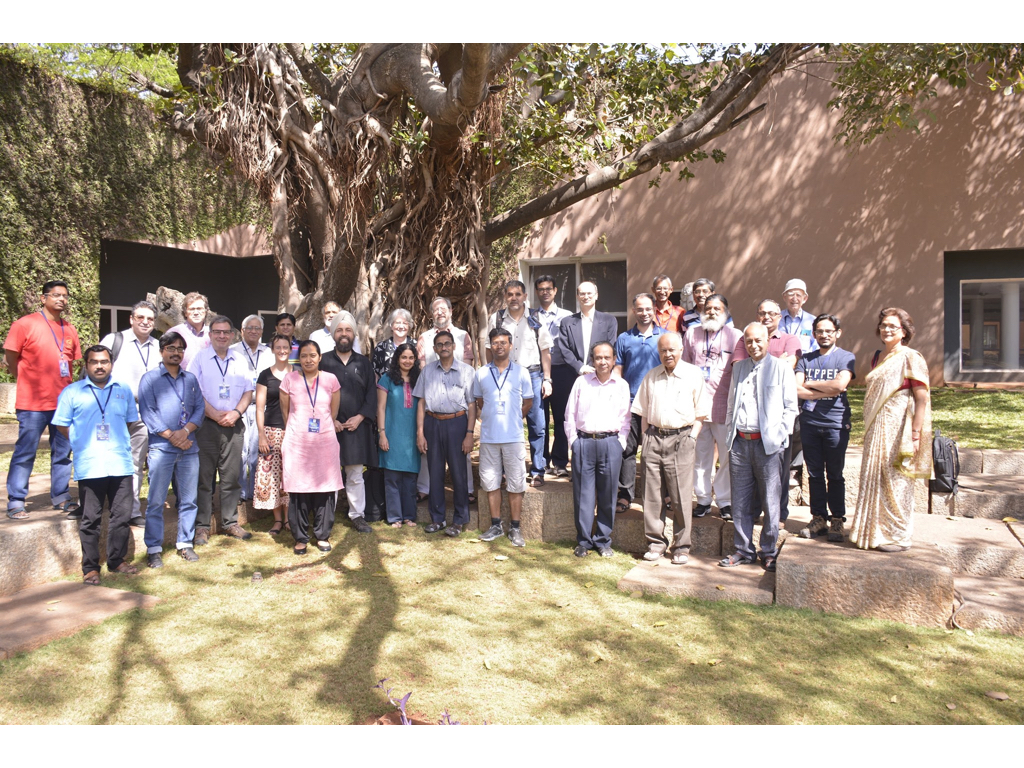}
\caption*{The 60th birthday conference, IUCAA, Pune, 2017} 
\end{figure*}
On the occasion of Paddy's fiftieth birthday in 2007, his students brought out a book `Vignettes in gravitation and cosmology' edited by Sriramkumar and Seshadri, and published by World Scientific. The publisher's blurb reads: 
"This book comprises expository articles on different aspects of gravitation and cosmology that are aimed at graduate students. The topics discussed are of contemporary interest assuming only an elementary introduction to gravitation and cosmology. The presentations are to a certain extent pedagogical in nature, and the material developed is not usually found in sufficient detail in recent textbooks in these areas." The book has detailed articles related to Paddy's diverse research interests, written by Jasjeet, Nissim, Tirthankar, Aseem, Seshadri, Shankaranarayanan, T. P. Singh, Sriramkumar and Sunu, where the authors reviewed a topic they were working on.

Paddy's sixtieth birthday conference `Aspects of gravity and cosmology' was held at IUCAA, Pune during March 9-11, 2017 and was a truly memorable occasion. Aseem and Tirthankar organised the meeting, which was very well attended by a large number of Paddy's students and collaborators and colleagues. A book `Gravity and the Quantum' was brought out on this occasion; it runs to nearly 500 pages and has twenty-five articles spanning astrophysics, cosmology, general relativity, and quantum gravity. It is edited by Jasjeet and Sunu, published by Springer, and the blurb reads "This book provides a compilation of in-depth articles and reviews on key topics within gravitation, cosmology and related issues. It is a celebratory volume dedicated to Prof. Thanu Padmanabhan ("Paddy"), the renowned relativist and cosmologist from IUCAA, India, on the occasion of his 60th birthday. The authors, many of them leaders of their fields, are all colleagues, collaborators and former students of Paddy, who have worked with him over a research career spanning more than four decades. Paddy is a scientist of diverse interests, who attaches great importance to teaching. With this in mind, the aim of this compilation is to provide an accessible pedagogic introduction to, and overview of, various important topics in cosmology, gravitation and astrophysics. As such it will be an invaluable resource for scientists, graduate students and also advanced undergraduates seeking to broaden their horizons.". The talks at the conference were recorded, and can be found at this link \url {https://youtube.com/playlist?list=PLGqalPsP5GRDW6CqUG7i_xUT-JUxU-qbq} This conference would be the last major get-together of Paddy and his academic family.

On September 25, 2021 IUCAA, Pune held a Paddy Memorial meeting, the recording of which is available at https://youtu.be/iZ2m1V0DIR4 A tribute page is at https://www.iucaa.in/tributes/paddy/

\section{The Paddy Academic Family Tree}
\noindent Paddy supervised seventeen graduate students, over a period of thirty-six years. They are T. R. Seshadri, Tejinder P. Singh, Jasjeet Singh Bagla, L. Sriramkumar, K. Srinivasan, Nissim Kanekar, S. Shankaranarayanan, Tirthankar Roy Choudhury, Sunu Engineer, Gaurang Mahajan, Sudipta Sarkar, Dawood Kothawala, Sanved V. Kolekar, Suprit Singh, Krishnamohan Parattu, Sumanta Chakraborty and  Karthik Rajeev. Aseem Paranjape, who was formally a student of T. P. Singh at TIFR,  worked with Paddy before, during and after his Ph.D. thesis days. Mohammad Sami had very close academic and personal ties with Paddy. Most of Paddy's students have had their own graduate students, and some of these grand-students now have their own students. So Paddy's academic family tree is currently in its fourth generation: you can find this academic family listed online here \url{https://academictree.org/physics/tree.php?pid=715864}.
\begin{figure*}[!h]
\centering
\includegraphics[width=12.00cm]{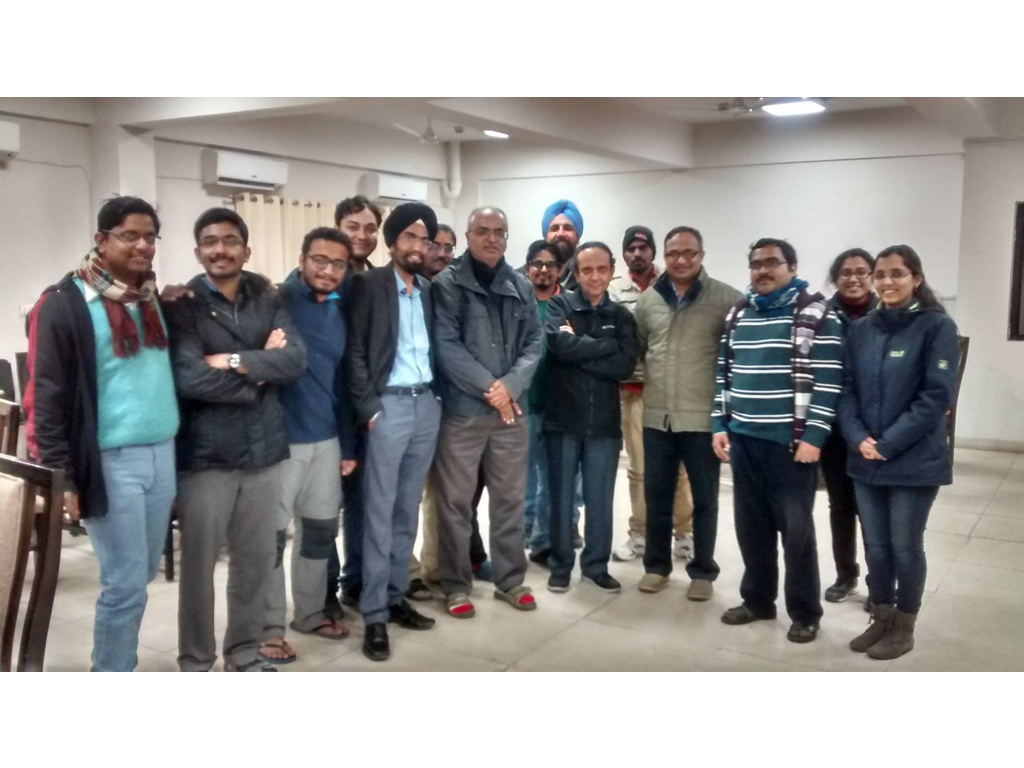}
\caption*{Paddy with several of his graduate students and a couple of grand-students at the 8th International Conference on Gravitation and Cosmology, IISER Mohali, 2015} 
\end{figure*}
Paddy also mentored several postdoctoral fellows during his career, starting with Urjit Yajnik and Seshadri Sridhar at TIFR. At IUCAA he mentored Sujoy Modak, Harvinder Kaur Jassal, Bibhas Ranjan Majhi, Valerio Faraoni, Kinjalk Lochan. J. Chan, Shiv Sethi, Ali Nayeri, Sourav Bhattacharya, Sandipan Sengupta and Krishnakanta Bhattacharya.

\medskip

\centerline{\it You are no longer with us, but you have left us with a task. To carry  your scientific legacy forward.}
\centerline{\it To that we commit, and then pass the baton on.}

\bibliography{biblioqmtstorsion}

\end{document}